\documentclass[preprint,3p,times]{elsarticle}
\usepackage{amssymb}
\usepackage[colorlinks,urlcolor=blue,linkcolor=blue,citecolor=black]{hyperref}
\usepackage{color,array}
\usepackage{amsmath}
\usepackage{amsfonts}
\usepackage{graphicx}
\usepackage{multirow}
\usepackage{booktabs}
\usepackage{threeparttable}
\usepackage{float}
\usepackage{gensymb}
\usepackage{subfig}
\usepackage{makecell}
\usepackage{bookmark}

\makeatletter
\newcommand*{\sectionbookmark}[1][]{%
  \bookmark[%
    level=section,%
    dest=\@currentHref,%
    #1%
  ]%
}
\makeatother

\usepackage{lineno}

\newfloat{figtab}{htb}{fgtb}
\makeatletter
  \newcommand\figcaption{\def\@captype{figure}\caption}
  \newcommand\tabcaption{\def\@captype{table}\caption}
\makeatother

\journal{Nature Communications}

\begin{document}

\begin{frontmatter}

\title{Integrating spoken instructions into flight trajectory prediction to optimize automation in air traffic control} 

\author[1]{Dongyue Guo}
\author[1]{Zheng Zhang}
\author[1]{Bo Yang}
\author[1]{Jianwei Zhang}
\author[1]{Hongyu Yang}
\author[1]{Yi Lin\corref{cor1}}
\ead{yilin@scu.edu.cn}
\cortext[cor1]{Corresponding author.}

\affiliation[1]{organization={College of Computer Science, Sichuan University}, 
            city={Chengdu},
            postcode={610000},
            country={China}}

\begin{abstract}
    The booming air transportation industry inevitably burdens air traffic controllers' workload, causing unexpected human factor-related incidents. 
    Current air traffic control systems fail to consider spoken instructions for traffic prediction, bringing significant challenges in detecting human errors during real-time traffic operations. 
    Here, we present an automation paradigm integrating controlling intent into the information processing loop through the spoken instruction-aware flight trajectory prediction framework.
    A 3-stage progressive multi-modal learning paradigm is proposed to address the modality gap between the trajectory and spoken instructions, as well as minimize the data requirements.  
    Experiments on a real-world dataset show the proposed framework achieves flight trajectory prediction with high predictability and timeliness, obtaining over 20\% relative reduction in mean deviation error. 
    Moreover, the generalizability of the proposed framework is also confirmed by various model architectures. 
    The proposed framework can formulate full-automated information processing in real-world air traffic applications, supporting human error detection and enhancing aviation safety. 
\end{abstract}


\end{frontmatter}
\section{Introduction} \label{sec1} 
In past decades, the air transportation industry is booming with increasing air traffic flow due to economic achievements. 
{{The workload of air traffic controllers (ATCos) is inevitably burdened by high traffic density, which causes great challenges in providing safe and efficient services for air traffic control (ATC). 
The increasing air traffic flow in recent years has led to a rise in aviation incidents caused by various human factors (e.g., cognitive workload and weakened situational awareness), bringing a huge risk to ATC safety.}}
Based on the investigation of \cite{KELLY2019155, XUE20181}, 70\% of aviation incidents are related to human factors, which inspires us to reconsider the protection of human factor-related risks in aviation fields.

In real-time ATC operation, most of the human factor-related risks are caused by errors in the ATC communication procedure \cite{anne}, such as unsafe ATC decisions, mishearing, and misunderstanding.  
The major limitation to detecting human factor-related risks is that current ATC systems fail to consider the human intention (represented by the controlling intent of the spoken instructions) for traffic prediction due to human-in-the-loop (HITL) natures. 
Specifically, in the ATC procedure, ATCos make decisions based on their awareness of the traffic dynamics and issue spoken instructions to negotiate with the pilots in ATC communication via very high-frequency (VHF) radiotelephony. 
Once the aircrews correctly readback the ATC instruction, they perform the required aircraft operations according to the controlling intent of the ATC instruction.
{{The flight will be in a high-maneuvering status (denoted as instruction-driven maneuvering flight scenarios in this work) with rapid and complex transition patterns of the flight trajectory.}} 

In this procedure, the most typical accidents by human errors can broadly categorized into two types: the unsafety ATC instructions issued by ATCos (within potential conflict) and incorrect operation of pilots (misunderstanding or failure to adhere to the provided ATC instructions) \cite{LIN2023366}. 
Notable examples include the Überlingen Mid-Air Collision in 2002 and the Haneda Airport runway collision in 2024, which resulted in significant losses and impacts for passengers, airlines, etc \cite{masys2005systemic, 404800142}. 
Despite enormous efforts devoted to enhancing safety measures in the ATC procedure \cite{zhang2020bayesian, 2017-4388, LinDCWZY20}, such incidents have not yet been resolved due to the aforementioned HITL nature, posing a continuous threat to people's lives and safety. 
{{In this context, predicting the influence of ATCo instructions on real-time traffic operations is a promising approach to detecting and protecting human error-related risks for ATC procedures. 
For example, i) if the issued ATC instruction contains potential risks, the predicted results can enable the downstream conflict detection applications to promptly identify possible flight conflicts;  
ii) if a pilot misunderstands or incorrectly executes an ATC instruction, the resulting flight trajectory will significantly deviate from the expections of ATC. 
Nevertheless, limited by the technical ability to tackle the spoken ATC instructions, the automation gap between human intention and ATC systems disables the development of automated measures for current ATC systems to detect human errors.}}

{{
Accurate flight trajectory prediction (FTP) results enable ATC participants to gather traffic dynamics in advance, supporting efficient decision-making and detecting potential conflicts in specific airspace regions \cite{53645, jiang2018svm, ChenG020}. 
Currently, short-term FTP serves as the vital application of traffic prediction in modern ATC systems, which aims to forecast the flight status of the aircraft in the future time instants \cite{9945661, aerospace9020091, ShiXP21, LIN2019105113}. 
Although existing FTP approaches can achieve the expected performance with constant transition patterns, such as the en-route phase, they are still facing great challenges in predicting the flight trajectory with maneuvering operations due to the intervention of human intentions.  
In general, the complicated maneuvering patterns caused by human intentions can be summarized below:}}
\begin{itemize}
    \item Microscopic maneuvering by pilots, including operational behaviors, real-time environmental factors, etc. 
    In \cite{Zhang_2023}, a sophisticated time-frequency analysis method was proposed to achieve the FTP task by implicitly capturing the complicated multi-scale maneuvering patterns, which harvest desired performance on complicated flight patterns.
    \item Macroscopic maneuvering by air traffic controllers, mainly concerning the spoken ATC instruction, which is the primary driving factor to cause maneuvering operations. 
    Under the aircraft separation rules in the ATC domain, macroscopic maneuvering is the most decisive factor in influencing flight trends, and is able to provide confident pre-warnings to flight conflicts. 
\end{itemize}

\begin{figure*}[b]
	\centering
	\includegraphics[width=0.98\textwidth]{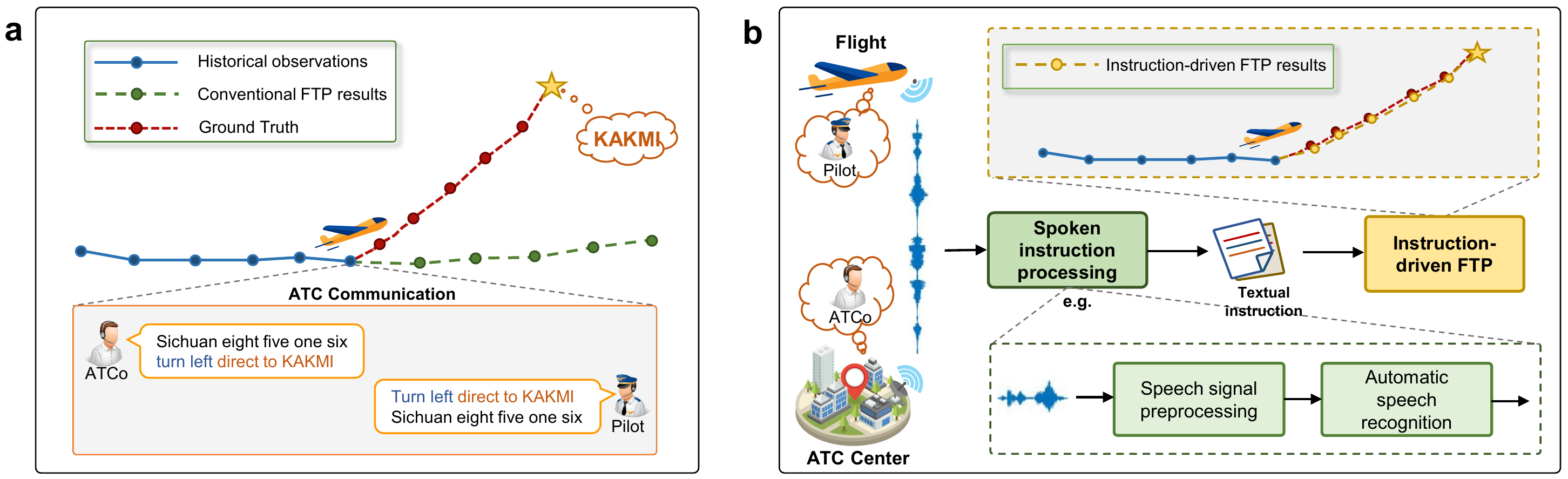}
	\caption{{{Comparison of conventional data-driven FTP and instruction-driven FTP. \textbf{a} An example of the ATC communication procedure and the challenges faced by the conventional FTP tools. \textbf{b} The logic flow of the proposed instruction-driven FTP paradigm.} }}
	\label{fig:1-1}
\end{figure*}

As illustrated in Figure \ref{fig:1-1}a, the controlling intent of real-time ATC spoken instructions (\textit{turn left direct to KAKMI}) serves as a driving factor to induce the macroscopic flight maneuvering operations. 
{{Since conventional data-driven FTP approaches typically only rely on historical trajectory observations and fail to consider read-time intents and required parameters, 
they have few trend-aware abilities and suffer from unreliable prediction results with delayed responses.}} 
Moreover, the flight trajectory can be seen as the most direct and ultimate manifestation of the controlling intent (ATC spoken instructions). 
Therefore, the ATC spoken instruction can be integrated into the FTP process to empower air traffic predictions, enabling the ATC systems to consider human factors automatically in a closed loop and further detect human errors. 
In this context, it is imperative to explicitly consider the ATC spoken instruction for the FTP task in a proper way. 

{{Inspired by this, in this work, an instruction-driven flight trajectory prediction paradigm is proposed to incorporate spoken-instruction into the automation process, including controlling intent understanding and resulting flight trajectory prediction.}} 
In this way, the most important information source (spoken instructions) of the human intention in real-time ATC operations can be perceived automatically with high timeliness, thereby enhancing the predictability of the ATCos' performance on traffic operations. 
Furthermore, the automation tools can be developed to detect potential human-related risks \cite{LinDCWZY20, LIN2023366}, and further enhance the safety and efficiency of traffic operations.

To be specific, we mainly focus on short-term FTP tasks within a few future time instants (approximately 1 to 10 minutes) based on historical observations and the spoken instructions, 
which further supports the downstream applications (e.g., conflict detection, and monitoring the ATC instruction performing processes). 
To this end, an intuitive approach is to develop multi-modal FTP approaches to explicitly consider both the spoken instructions and historical trajectory observations during the instruction-driven maneuvering flight scenarios. 
{{Unlike the concept in human/vehicle trajectory prediction domain \cite{8500493, 0003LFL21, DendorferOL20, Luo0DY20, 0003LCL23}, the term "multi-modal" in this work refers specifically to the fusion of data from different modalities through multi-modal learning \cite{3545572}.}} 
However, it is difficult to incorporate the spoken instructions into the FTP process due to the following challenges: 

\begin{itemize}
    \item {{In the ATC procedure, the communications between the ATCos and pilots are based on speech communication via VHF radiotelephony, while the flight trajectories are collected via binary structures and decoded into the modality of spatial-temporal data. 
    It is clear that the distinct modality gap between trajectory and spoken instructions brings great technical challenges to considering spoken instruction in FTP tasks.}} 

    \item An expected limitation of data-driven multi-modal approaches is the requirement for well-resourced trajectory-instruction pairs in the training process. 
    However, the collection, preprocessing, and annotation of trajectory-instruction paired samples are both time-consuming and labor-intensive. 
    {{It is challenging to train a multi-modal FTP model utilizing limited trajectory-instruction pairs.}} 
\end{itemize} 

Considering the abovementioned challenges, in this paper, a spoken instruction-aware flight trajectory prediction framework, called SIA-FTP, is innovatively proposed to implement instruction-driven FTP task, which further incorporates human intentions into an ATC automation process. 
{{In general, spoken instruction is the speech signal with considerable information redundancy, such as radiotelephony noise and speaker identity, which makes it challenging to incorporate the speech signal directly into the FTP task. 
Fortunately, based on our previous works on the Automatic Speech Recognition (ASR) technique in the ATC domain \cite{9174746, LinYLGZCZ21, aerospace8110348, abs211102041, 10106480}, 
the spoken instruction can be translated into high-confidence human/computer-readable transcripts, indicating the controlling intents and required detailed parameters.}}
Benefiting from the previous efforts, as depicted in Figure \ref{fig:1-1}b, the spoken instruction can be transcribed into textual modality by the existing ASR systems, which can further reduce the modality gap. 
In this context, the primary challenge of the SIA-FTP framework is to incorporate the textual spoken instructions into the FTP model under limited trajectory-instruction paired samples. 
It is expected that the proposed SIA-FTP framework can leverage the complementary and diverse information from both textual spoken instructions and spatial-temporal trajectory modalities to enhance the performance of FTP tasks.

In practice, compared to the limited paired data, the unimodal trajectory and text instruction data are well-resourced and can be easily obtained separately, as in our previous work \cite{9945661, LinDCWZY20}. 
Therefore, in this paper, a 3-stage progressive multi-modal learning paradigm is designed to train the proposed SIA-FTP framework, including trajectory-based FTP pre-training, intent-oriented instruction embedding learning, and multi-modal FTP fine-tuning, as described below:
\begin{itemize}
    \item  Stage 1: trajectory-based FTP pre-training stage, a multi-horizon FTP model proposed in our previous work, named FlightBERT++ \cite{guo2023flightbert++}, is applied to learn the spatial-temporal movement patterns only by trajectory data samples, in which the temporal trajectory prediction (predicting future trajectory only based on historical observations) serves as the pre-training task.

    \item  Stage 2: intent-oriented instruction embedding learning stage, a BERT-based architecture is firstly introduced to learn abstract compact text representations, followed by a multi-label intent identification (IID) task to learn the discriminative embeddings among different controlling intents from the text instructions.

    \item  Stage 3: multi-modal FTP finetuning stage, a simple yet effective modal fusion strategy is designed to explicitly bridge the pre-trained FTP model and intent identification model to conduct multi-modal FTP model, which can incorporate the instruction embedding into the pre-trained FTP model.
\end{itemize}

Finally, the SIA-FTP model is trained on the limited trajectory-instruction pairs to finetune the model parameters, which is expected to learn the flight transition patterns considering specific instructions (intent and required parameters). 
Thanks to the FlightBERT++ framework with the explicit fusion of controlling intents, the proposed SIA-FTP model is able to predict future multi-horizon trajectory points in a non-autoregressive manner, 
as well as the macroscopic maneuvering awareness, which has high performance and efficiency to enhance real-world applicability.  

To validate the proposed SIA-FTP framework, a real-world dataset is built to conduct the experiments, which are collected from the industrial ATC systems in China. 
The experimental results demonstrate that the proposed SIA-FTP framework achieves impressive results in instruction-driven high-maneuvering flight processes, achieving over 20\% relative reduction of the mean deviation error across 15 prediction horizons (5 minutes) compared to the best baseline. 
All the proposed techniques and strategies are confirmed to provide the desired performance improvement. 
Most importantly, extensive visualizations and in-depth diagnostic studies are conducted to enhance the interpretability and generalizability of the proposed SIA-FTP framework.
It is believed that the proposed framework can be a promising approach to empower FTP-based downstream applications, especially for conflict detection and resolution, and erroneous ATC instruction identification, thereby enhancing the safety and efficiency of ATC operations. 
In summary, this work contributes to the human error detection in real-time ATC operations and resulting instruction-driven FTP tasks in the following ways:  
\begin{itemize}
    \item This work innovatively defines the instruction-driven FTP task (i.e., incorporating spoken instruction into the FTP process), which has solid significance and applicability to the ATC work.
    In addition, the proposed framework can incorporate textual instructions into FTP tasks to improve the model performance in instruction-driven maneuvering scenarios. 
    
    \item A 3-stage progressive multi-modal learning paradigm is designed to develop the proposed SIA-FTP framework, which enables the model to achieve the desired performance under the limited trajectory-instruction paired samples. 
    
    \item A multi-label intent identification method is proposed to understand controlling intents from text instructions, which extracts informative intent-oriented instruction embeddings and projects them into a compact embedding space to support modal fusion during joint optimization.  
    
    \item A simple yet effective multi-modal fusion mechanism is designed to fuse the trajectory embedding and the intent-oriented instruction embedding, thereby supporting spoken instruction awareness in the FTP process. 
    
    \item A trajectory-instruction dataset is built to conduct the experiments, which can be regarded as benchmarking in future human intention automation studies.  
    Extensive experiments demonstrated the efficiency and effectiveness of the proposed SIA-FTP framework. 
\end{itemize}

\section{Results} \label{sec3}
\subsection{Task Overview} \label{sec3.1}
In general, the short-term FTP tasks can be formulated the spatial-temporal sequential modeling problems, which predict flight status in a few minutes based on the observation sequence. 
Let the observation sequence be $O_{t-k+1:t} = \{p_{t-k+1}, ..., p_{t-1}, p_t\}$, the FTP model aims to forecast the $P_{t+1:t+n} = \{p_{t+1}, p_{t+2}, ..., p_{t+n}\}$ based on the past $k$ observations $O_{t-k+1:t}$, where the $p_t$ represents a trajectory point in time step $t$. 
The conventional FTP tasks can be mathematically described as follows: 
\begin{equation} \label{eq1}
    P_{t+1:t+n} = \{p_{t+1}, p_{t+2}, ..., p_{t+n}\} =  \mathcal{F} (O_{t-k+1:t})
\end{equation}
\noindent where the $n, k$ are the number of prediction trajectory points and the observed sequence length, respectively. 
$\mathcal{F}(\cdot)$ denotes the learnable FTP model. 

In this work, the SIA-FTP tasks can be defined as Eq. (\ref{eq2}), where the $\mathrm{SI}$ is the textual spoken instruction of ATCo that is confirmed by the pilot in time step $t$. 
\begin{equation} \label{eq2}
    P_{t+1:t+n} = \{p_{t+1}, p_{t+2}, ..., p_{t+n}\} =  \mathcal{F} (O_{t-k+1:t}, \mathrm{SI}) 
\end{equation}

Additionally, the definition of the trajectory point $p_t$ is presented in Eq. (\ref{eq3}). 
Specifically, a total of six trajectory attributes that describe the flight status are selected to form the $p_t$, including longitude ($\mathrm{Lon}$), latitude ($\mathrm{Lat}$), altitude ($\mathrm{Alt}$), and corresponding velocities ($\mathrm{Vx}, \mathrm{Vy}, \mathrm{Vz}$) along these dimensions. 
In this work, the Lon, Lat, and Alt serve as the primary attributes of the four-dimensional (4D) FTP task, while the velocities are employed as the auxiliary attributes. 
\begin{equation} \label{eq3}
    p_t = [\mathrm{Lon}_t, \mathrm{Lat}_t, \mathrm{Alt}_t, \mathrm{Vx}_t, \mathrm{Vy}_t, \mathrm{Vz}_t]
\end{equation}

\subsection{Dataset and Data Preprocessing} \label{sec3.2}
To validate the effectiveness of the proposed framework, a multi-modal situational dataset M2ATS \cite{3613759} is used to train the proposed SIA-FTP framework, which is collected from the real-world ATC system in China, from February 19 to February 27, 2021. 
The M2ATS covers the diverse ATC and flight operation data, including the flight trajectories data, flight plans, airspace information, and speech of the ATC communication with golden labels. 
In this work, we divided the M2ATS into three subsets to support the experiments of the SIA-FTP framework, i.e., trajectory subset, text instruction subset, and trajectory-instruction subset. 
The data split strategy of train, validation, and test set follows the original M2ATS. 
Specifically, the data of the first 7 days serve as training data, whereas the data from the rest two days are applied to the validation and test, respectively. 
It is noted that the above data split strategy is applied to all the experiments and training phases in this work. 
The detailed data preprocessing process is described in the Supplementary Information (Section Data Preprocessing).

\subsection{Comparison Baselines} \label{sec3.3}
In this work, a total of 7 competitive approaches serve as baselines. 
According to the inference style of the multi-horizon prediction process, the baselines are further categorized into iterative prediction models (LSTM, Transformer, Kalman-Filter, FlightBERT, and WTFTP) and direct prediction models (LSTM+Attention, and FlightBERT++). 
The iterative prediction models perform only one-step prediction in an inference procedure, and the predicted results will serve as pseudo observations to obtain the multi-horizon results iteratively.  
In contrast, the direct prediction models can generate multi-horizon prediction results through one-pass inference. 
The description of the baseline models is listed as follows. 

\begin{itemize}
    \item LSTM: The LSTM network is applied to build the FTP model \cite{ShiXPYZ18}, which is typically used for sequence modeling in many tasks. 
    \item Transformer: A vanilla Transformer architecture is adapted to develop the FTP model \cite{VaswaniSPUJGKP17}. 
    \item Kalman-Filter: A typical model-driven flight state estimation algorithm based on historical observations and generally applied in target tracking scenarios. 
    In this work, the Kalman-Filter (KF) is employed to perform the FTP task \cite{1960A}. 
    \item FlightBERT: An FTP framework proposed in our previous work, which innovatively converts the FTP tasks into a multi-binary classification paradigm \cite{9945661}. 
    \item WTFTP: A time-frequency analysis based FTP framework proposed in our previous work, which has demonstrated that the WTFTP is robust to the high-maneuvering flight scenarios \cite{Zhang_2023}. 
    \item LSTM+Attention: A Seq2Seq architecture to predict the flight state of multiple future time steps directly. 
    In this experiment, an Attention-based Encoder-Decoder LSTM network \cite{BahdanauCB14} is adapted to perform the FTP task.  
    \item FlightBERT++: As described in Section Methods, the FlightBERT++ is a non-autoregressive multi-horizon FTP framework that is also employed as the pre-trained FTP model in the proposed SIA-FTP framework \cite{guo2023flightbert++}. 
\end{itemize}

\subsection{Experimental Configuration} \label{sec3.4}
In the LSTM, Transformer, WTFTP, and LSTM+Attention baselines, the z-score normalization algorithm is applied to process the value into [0,1] for longitude and latitude attributes due to the sparse specificities of their distributions, 
while the other attributes are normalized into [0,1] through the max-min algorithm. 
Additionally, we employ the MSE loss function to train these models on the trajectory subset. 
For the BE-based baselines, the experimental configurations follow our original works \cite{9945661, guo2023flightbert++}. 

In the SIA-FTP framework, all the configurations in the trajectory-baed FTP pre-training stage are the same as the FlightBERT++. 
For the intent-oriented instruction embedding learning stage, the Chinese characters and English words serve as the basic units (tokens) for the language modeling.
A 256-dimensional hidden state is set to pre-train the BERT model. 
After the unsupervised text representation learning, an FC layer with 16 neurons is used as the prediction head to output the probabilities of the intent class. 
In the multi-modal FTP finetuning stage, the hidden state dimension for linear layers in the MLP is set to 1024 and 512, respectively. 

In this work, the Mean Absolute Error (MAE), Mean Absolute Percentage Error (MAPE), Root of Mean Square Error (RMSE) and Mean Deviation Error (MDE) are applied to evaluate the model performance and baselines. 
Among them, the MAE, MAPE, and RMSE are the common metrics to evaluate each attribute of trajectory point separately, while the MDE is proposed to measure the distance (nautical miles (NM)) of trajectory points between predictions and ground truth in three-dimensional (3D) airspace. 
In this way, it is believed that the performance of the proposed model and baselines can be comprehensively evaluated.
The detailed definition of the abovementioned metrics can be found in Supplementary Information (Section Definition of Evaluation Metrics).

The Adam optimizer with $10^{-4}$ initial learning rate is applied to train all the above deep learning-based models. 
In this work, all the experiments are implemented with the open-source deep learning framework PyTorch 1.9.0. 
The models are trained on the server configured with Ubuntu 16.04 operating system, 8*NVIDIA GeForce RTX 2080 GPU, Intel(R) Core(TM) i7-7820X@3.6GHz CPU, and 128 GB memory.

\subsection{Results and Quantitative Analysis} \label{sec3.6}

\begin{table}[t]
     \footnotesize
    \centering
    \setlength\tabcolsep{5 pt} 
    \caption{{{The experimental results of the proposed framework and baselines. The bold ones denote the best performance on the corresponding metric.}} } \label{tab2}
    \begin{tabular}{c|c|c|ccc|ccc|ccc|c}
    \toprule
    \multirow{2}{*}{\textbf{Style}} & \multirow{2}{*}{\textbf{Methods}} & \multirow{2}{*}{\textbf{Horizon}} & \multicolumn{3}{c|}{\textbf{MAE} $\downarrow$} & \multicolumn{3}{c|}{\textbf{MAPE (\%)} $\downarrow$} & \multicolumn{3}{c|}{\textbf{RMSE} $\downarrow$} & \multirow{2}{*}{\textbf{MDE} $\downarrow$} \\ \cline{4-12}
                                    &                                   &                                   & Lon        & Lat       & Alt      & Lon          & Lat          & Alt       & Lon        & Lat        & Alt      &                               \\ \midrule
    \multirow{16}{*}{Iterative}     & \multirow{4}{*}{LSTM}             & 1                                 & 0.0050     & 0.0055    & 2.15     & 0.0046       & 0.0205       & 0.24      & 0.0071     & 0.0078     & 3.24     & 0.48                          \\
                                    &                                   & 3                                 & 0.0063     & 0.0066    & 4.37     & 0.0058       & 0.0246       & 0.48      & 0.0093     & 0.0097     & 7.36     & 0.58                          \\
                                    &                                   & 9                                 & 0.0138     & 0.0135    & 13.41    & 0.0128       & 0.0502       & 1.47      & 0.0251     & 0.0246     & 21.38    & 1.23                          \\
                                    &                                   & 15                                & 0.0253     & 0.0260    & 23.63    & 0.0234       & 0.0969       & 2.57      & 0.0479     & 0.0587     & 37.80    & 2.31                          \\ \cline{2-13} 
                                    & \multirow{4}{*}{Transformer}      & 1                                 & 0.0032     & 0.0041    & 1.69     & 0.0030       & 0.0150       & 0.19      & 0.0044     & 0.0051     & 2.82     & 0.33                          \\
                                    &                                   & 3                                 & 0.0065     & 0.0082    & 4.10     & 0.0060       & 0.0305       & 0.45      & 0.0095     & 0.0111     & 7.35     & 0.66                          \\
                                    &                                   & 9                                 & 0.0172     & 0.0208    & 12.12    & 0.0160       & 0.0771       & 1.32      & 0.0282     & 0.0309     & 21.71    & 1.70                          \\
                                    &                                   & 15                                & 0.0292     & 0.0321    & 24.59    & 0.0271       & 0.1192       & 2.70      & 0.0505     & 0.0611     & 41.51    & 2.72                          \\ \cline{2-13} 
                                    & \multirow{4}{*}{Kalman-Filter}    & 1                                 & 0.0044     & 0.0047    & 3.40     & 0.0041       & 0.0177       & 0.38      & 0.0165     & 0.0087     & 5.53     & 0.41                          \\
                                    &                                   & 3                                 & 0.0084     & 0.0089    & 7.15     & 0.0077       & 0.0331       & 0.79      & 0.0296     & 0.0164     & 11.79    & 0.77                          \\
                                    &                                   & 9                                 & 0.0266     & 0.0250    & 19.87    & 0.0247       & 0.0928       & 2.21      & 0.0871     & 0.0462     & 32.12    & 2.29                          \\
                                    &                                   & 15                                & 0.0639     & 0.0592    & 37.53    & 0.0594       & 0.2199       & 4.15      & 0.2512     & 0.1208     & 74.96    & 5.50                         \\ \cline{2-13} 
                                    & \multirow{4}{*}{FlightBERT}       & 1                                 & 0.0029     & 0.0018    & 1.63     & 0.0027       & 0.0068       & 0.18      & 0.0235     & 0.0311     & 12.33    & 0.25                          \\
                                    &                                   & 3                                 & 0.0047     & 0.0032    & 3.54     & 0.0043       & 0.0121       & 0.39      & 0.0309     & 0.0181     & 16.33    & 0.39                          \\
                                    &                                   & 9                                 & 0.0136     & 0.0100    & 12.51    & 0.0126       & 0.0373       & 1.39      & 0.0381     & 0.0286     & 28.83    & 1.13                          \\
                                    &                                   & 15                                & 0.0287     & 0.0192    & 34.18    & 0.0267       & 0.0715       & 3.78      & 0.0733     & 0.0430     & 55.60    & 2.24                          \\ \cline{2-13}
                                    & \multirow{4}{*}{WTFTP}            & 1                                 & 0.0019     & 0.0016    & 1.10     & 0.0017       & 0.0060       & 0.12      & 0.0031     & \bf{0.0025}& \bf{2.08}& 0.16                          \\
                                    &                                   & 3                                 & \bf{0.0032}& 0.0030    & 5.02     & 0.0029       & 0.0112       & 0.56      & \bf{0.0052}& 0.0054     & 8.77     & \bf{0.28}                         \\
                                    &                                   & 9                                 & 0.0136     & 0.0129    & 19.07    & 0.0127       & 0.0481       & 2.10      & 0.0247     & 0.0253     & 30.32    & 1.21                          \\
                                    &                                   & 15                                & 0.0260     & 0.0240    & 27.64    & 0.0241       & 0.0897       & 3.07      & 0.0492     & 0.0455     & 41.67    & 2.25                          \\   \midrule
        \multirow{12}{*}{Direct}    & \multirow{4}{*}{LSTM+Attention}   & 1                                 & 0.0059     & 0.0060    & 2.19     & 0.0054       & 0.0225       & 0.25      & 0.0084     & 0.0091     & 2.98     & 0.53                          \\
                                    &                                   & 3                                 & 0.0058     & 0.0057    & 4.21     & 0.0054       & 0.0216       & 0.47      & 0.0082     & 0.0086     & 7.19     & 0.52                          \\
                                    &                                   & 9                                 & 0.0124     & 0.0115    & 14.41    & 0.0115       & 0.0428       & 1.58      & 0.0243     & 0.0235     & 25.42    & 1.09                          \\
                                    &                                   & 15                                & 0.0201     & 0.0176    & 20.07    & 0.0186       & 0.0658       & 2.19      & 0.0425     & 0.0406     & 35.67    & 1.71                          \\ \cline{2-13} 
                                    & \multirow{4}{*}{FlightBERT++}     & 1                                 & \bf{0.0018}& \bf{0.0014} & 1.02   & 0.0017       & 0.0055       & 0.12      & 0.0033     & 0.0027     & 2.94     & 0.16                          \\
                                    &                                   & 3                                 & 0.0037     & 0.0032    & 4.95     & 0.0035       & 0.0122       & 0.56      & 0.0064     & 0.0061     & 9.22     & 0.33                          \\
                                    &                                   & 9                                 & 0.0120     & 0.0105    & 13.80    & 0.0112       & 0.0392       & 1.53      & 0.0236     & 0.0203     & 25.14    & 1.03                          \\
                                    &                                   & 15                                & 0.0180     & 0.0163    & 17.37    & 0.0167       & 0.0609       & 1.92      & 0.0348     & 0.0309     & 33.53    & 1.55                          \\ \cline{2-13} 
                                    & \multirow{4}{*}{SIA-FTP}          & 1                                 & \bf{0.0018}  & \bf{0.0014}    & \bf{0.85}     & \bf{0.0016}  & \bf{0.0053}       & \bf{0.10}      & \bf{0.0030}     & 0.0027          & 2.57          & \bf{0.15}                          \\
                                    &                                   & 3                                 & 0.0034       & \bf{0.0028}    & \bf{3.42}     & \bf{0.0032}  & \bf{0.0105}       & \bf{0.38}      & 0.0061          & \bf{0.0053}     & \bf{6.91}     & 0.29                               \\
                                    &                                   & 9                                 & \bf{0.0089}  & \bf{0.0078}    & \bf{7.51}     & \bf{0.0083}  & \bf{0.0290}       & \bf{0.83}      & \bf{0.0168}     & \bf{0.0160}     & \bf{14.53}    & \bf{0.77}                          \\
                                    &                                   & 15                                & \bf{0.0141}  & \bf{0.0124}    & \bf{9.75}     & \bf{0.0131}  & \bf{0.0464}       & \bf{1.07}      & \bf{0.0285}     & \bf{0.0253}     & \bf{19.30}    & \bf{1.22}                          \\ \bottomrule
    \end{tabular}
    \begin{tablenotes}
        \item[1] $\downarrow$ represents minimization indicators.
        \item[2] The bold ones denote the best performance on the corresponding metric.
    \end{tablenotes}
\end{table}

The experimental results are reported in Table \ref{tab2}. 
In general, thanks to the maneuvering ATC instruction incorporation, the proposed SIA-FTP framework achieves the desired performance in terms of all the four proposed evaluation metrics, for all prediction horizons. 
In the $1^{st}$ prediction horizon, the baseline FlightBERT++ achieves comparable performance with the proposed SIA-FTP, in terms of the MAE of the $Lon$ dimension and the $Lat$ dimension. 
The results can be attributed that in the first prediction horizon, the instruction-driven factors only have a minor influence on the flight trajectory, and the FlightBERT++ is the trajectory representation learning model of the SIA-FTP, 
therefore, it can obtain some comparable indicators during the prediction process.  
To be specific, by analyzing the samples from the dataset, it is found that some flights might have not executed the instructions in the $1^{st}$ prediction horizon (20 seconds) due to the operational habits of the pilot. 

As can be seen, the proposed SIA-FTP framework is superior to all the baselines among the 4 evaluation metrics in the multi-horizon prediction results, even obtaining over 20\% MDE reduction in the 9- and 15-horizons compared to the FlightBERT++. 
In general, thanks to the non-autoregressive prediction mechanism of the FlightBERT++, the SIA-FTP achieves higher multi-horizon prediction performance. 
Most importantly, the SIA-FTP is able to be aware of the controlling intent by integrating the spoken instruction into the FTP model explicitly. 
For the competitive data-driven baseline WTFTP approach, as demonstrated in \cite{Zhang_2023}, thanks to the capability of time-frequency analysis and in-depth feature extraction, 
it can effectively capture microscopic maneuvering patterns of trajectory implicitly and obtain comparable performance against the proposed SIA-FTP framework within 3-horizons. 
However, as the prediction horizon increases and the flight enters the macroscopic maneuvering phase, it is difficult to estimate the flight intent implicitly in longer horizons, which results in performance degradation in multi-horizon predictions. 
{{The above experimental results demonstrate that, by considering the real-time spoken instructions, the proposed SIA-FTP framework is a promising solution to improve the prediction precision for FTP tasks in the ATC domain, especially for the instruction-driven high macroscopic maneuvering scenes.}} 

Among the selective baselines, it can be found that the iterative multi-horizon prediction approaches are prone to suffering from error accumulation with the increase of the prediction steps. 
Taking the Kalman-Filter approach as an example, it loses the history observation to update the parameters of the system equation in the multi-horizon prediction process, leading to a significant degradation in performance. 
In contrast, benefiting from the multi-horizon prediction strategy, the direct multi-horizon approaches facilitate capturing the global trend and are superior to the iterative approaches.   
It can be seen from the results that, except for the proposed SIA-FTP framework, the FlightBERT++ achieves superior performance over other baselines. 
This also supports our motivation to employ FlightBERT++ for trajectory-based FTP pre-training in the proposed SIA-FTP framework. 

\begin{figure*}[htbp] 
    \centering
    \includegraphics[width=\textwidth]{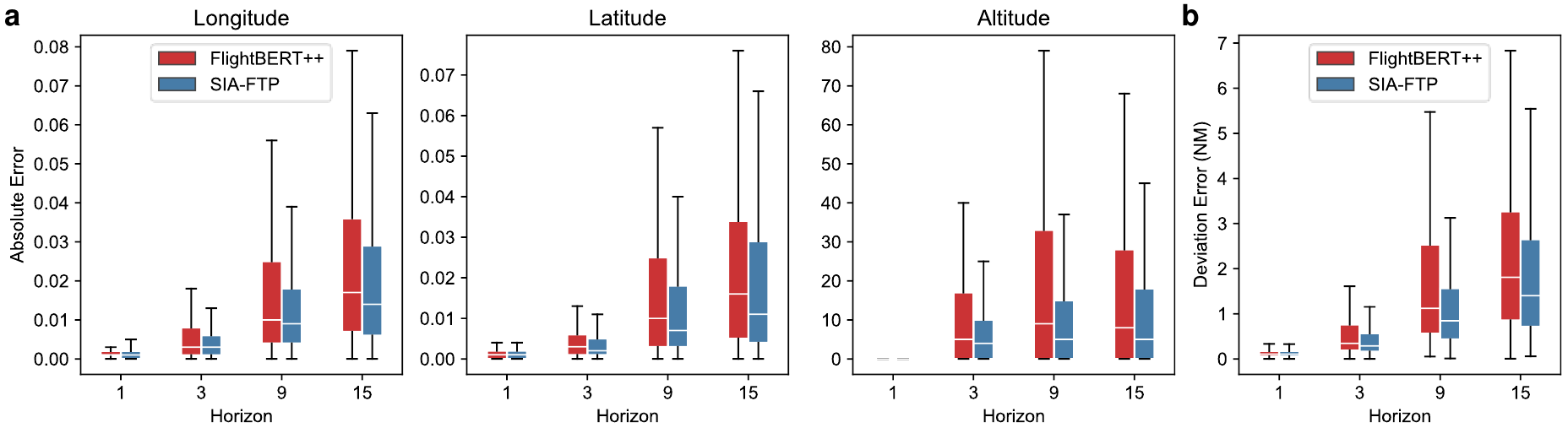}
    \caption{Error distribution of the FlightBERT++ and the proposed SIA-FTP. 
    \textbf{a} The distribution of absolute error for Longitude, Latitude, and Altitude attributes across different prediction horizons. \textbf{b} The distribution of deviation error across different prediction horizons.
    Boxplots of a and b show the median (center line), and 1st and 3rd quartiles (Q1 and Q3, respectively). 
    The error bars correspond to the Q1-(1.5*IQR) and Q3 + (1.5*IQR) range (IQR = Inter-Quartile Range). 
    Data points below Q1 – (1.5*IQR) or above Q3 + (1.5*IQR) are considered outliers and not shown in the boxplots.
    Source data are provided as a Source Data file.}\label{fig_err_dis}
\end{figure*} 

In addition, Figure \ref{fig_err_dis} illustrates the error distribution of the FlightBERT++ and the proposed SIA-FTP framework across $1^{st}, 3^{rd}, 9^{th}$ and $15^{th}$ prediction horizons. 
Figure \ref{fig_err_dis}a presents the absolute error for longitude, latitude, and altitude attributes, it is evident from the box plots that the proposed SIA-FTP framework consistently achieves lower absolute errors across all horizons compared to FlightBERT++. 
Notably, the proposed SIA-FTP harvests higher error reductions at longer horizons (9 and 15), where the ATC instructions provide stronger driven influences on flight patterns. 
The distribution of deviation error (DE) across different prediction horizons is reported in Figure \ref{fig_err_dis}b. 
The results show that SIA-FTP maintains a lower deviation error across all horizons, further reinforcing its effectiveness by incorporating ATC instructions. 
{{Furthermore, the error distribution based on absolute percentage error and squared error, as shown in Supplementary Figure 1, further corroborates the above conclusions.
Additionally, the statistical analysis across four metrics also confirms that SIA-FTP significantly outperforms FlightBERT++ in predicting flight trajectories over longer horizons, which can be found in Supplementary Information (Supplementary Table 1).  
The significant error reduction achieved by SIA-FTP confirms its potential for enhancing the performance of FTP in instruction-driven maneuvering flight scenarios.}}

\subsection{Visualization and Qualitative Analysis} \label{sec_vis}
\begin{figure*}[htbp] 
    \centering
    \includegraphics[width=0.95 \textwidth]{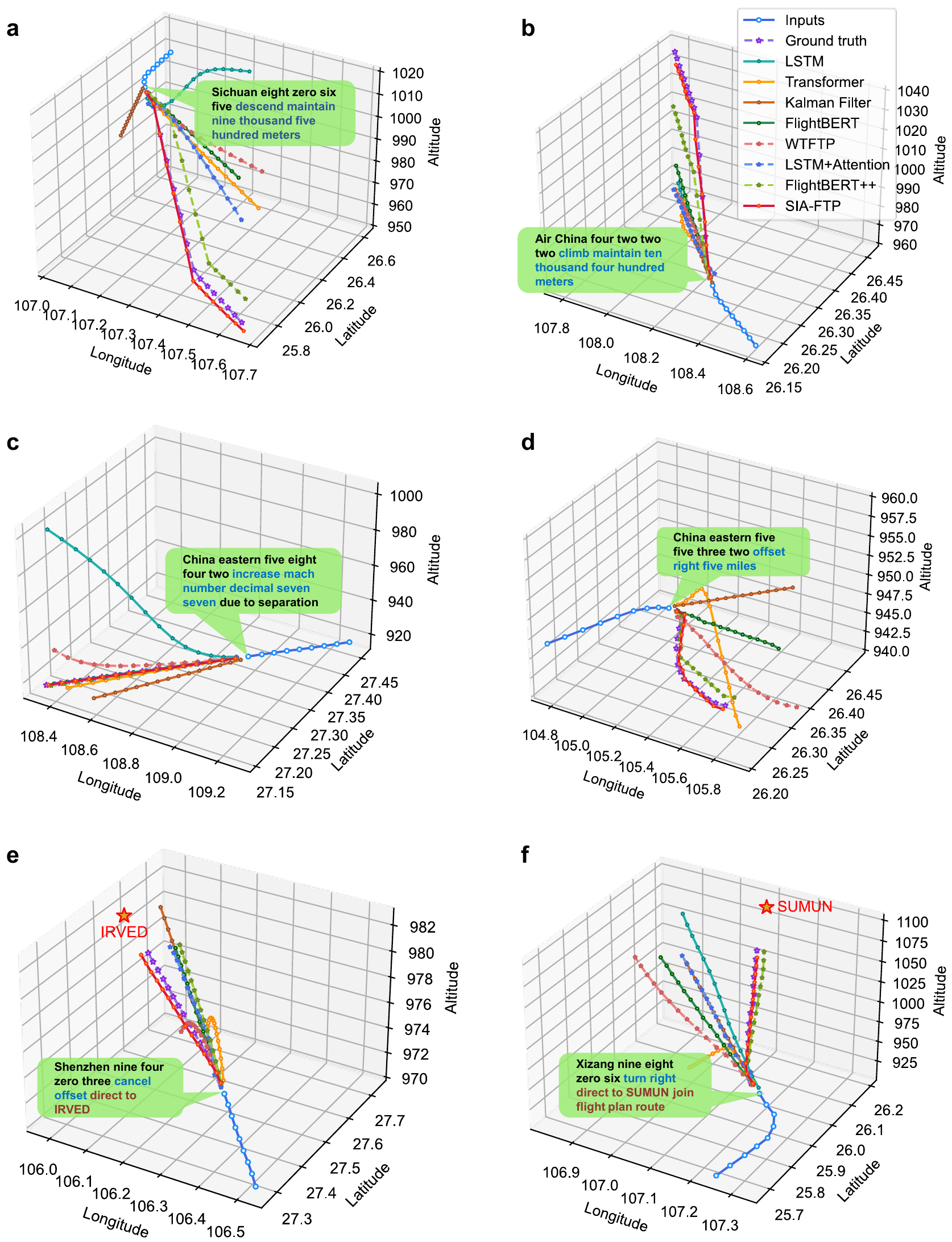} 
    \caption{Visualization of the trajectory prediction results with different maneuvering controlling intents. 
    \textbf{a-b} ALT\_ADJ.  \textbf{c} SPD\_ADJ. \textbf{d} OFFSET. \textbf{e} CANOFF\&FLYTO. \textbf{f} HEAD\_ADJ\&FLYTO. 
    The ATC instructions are presented in the chatbox, where the color fonts indicate the keywords of the controlling intents. 
    Source data are provided as a Source Data file.}\label{fig_vis6}
\end{figure*}

To vividly demonstrate the effectiveness of the SIA-FTP framework across various maneuvering intents, in this section, a total of six representative flight trajectories with ATC instructions are selected to visualize the predictive results of each model. 
The prediction results of the proposed framework and baselines are depicted in Figure \ref{fig_vis6}, in which the blue lines indicate the model inputs (i.e., observations of 9 trajectory points), and the altitude is measured in units of 10 meters.
It is important to note that certain inaccurate predictions made by baselines are removed from Figure \ref{fig_vis6} to enhance readability. 

As shown in Figure \ref{fig_vis6}, it is evident from the visualization results that the proposed SIA-FTP method has the ability to perform consistent predictions under the instruction-driven maneuvering scenarios for all six controlling intents in a multi-horizon manner. 
Figure \ref{fig_vis6}a and Figure \ref{fig_vis6}b illustrate the prediction results of two different ALT\_ADJ (altitude adjustment) ATC instructions, including descending and climbing flight phases, respectively. 
It can be seen from the visualizations that almost all baselines fail to predict the flight intents in future horizons when the flights are under complex ATC instructions. 
As demonstrated in our previous work \cite{guo2023flightbert++}, the FlightBERT++ model has certain capabilities to predict future flight intent, which benefits from learning the flight transition patterns from the volumes of historical trajectory data. 
As stated in the WTFTP, the multi-horizon (over 9 steps) prediction is a limitation to improving the FTP performance, which also supports the multi-horizon prediction mechanism in this work.
For instance, in Figure \ref{fig_vis6}a and Figure \ref{fig_vis6}b, the FlightBERT++ implicitly predicts the climbing/descending action to support the FTP task.  
However, FlightBERT++ falls short in delivering desired prediction results due to the inability to explicitly consider the detailed controlling intent from real-time ATC instructions. 
For instance, in Figure \ref{fig_vis6}b, the predicted flight trajectory of the FlightBERT++ only reaches the altitude of approximately 10,100 meters, rather than the specified 10,400 meters (target altitude) mentioned in the ATC instruction. 
{{Fortunately, thanks to the explicit fusion of the spoken instruction, the proposed SIA-FTP framework not only provides timely aircraft maneuvering predictions but also exhibits detailed awareness of the "maintain" intent from ATC instructions for future horizons after "climbing/descending" to the target altitude.}}

Similarly, the FlightBERT++ accurately predicts the flight intent of "OFFSET" in Figure \ref{fig_vis6}d but fails to exactly estimate the number of miles for the offset maneuver. 
In practice, the specific parameters of the controlling intents (such as altitude, speed, and miles of the offset) are typically determined by ATCos according to the real-time ATC situations. 
Thus, it is difficult for the trajectory-based FTP models to learn these patterns from the historical data and provide accurate prediction results, that is why we consider the explicit fusion of the controlling intents and required parameters. 

In addition, the prediction results presented in Figure \ref{fig_vis6}e and Figure \ref{fig_vis6}f demonstrate that: i) the proposed SIA-FTP framework performs accurate predictions even with multiple controlling intents in an ATC instruction; 
ii) external airspace knowledge, such as the position of waypoints (IRVED, SUMUN), and projections of the direction (turn right/left), is learned from historical trajectory samples in the multi-modal FTP finetuning process. 
The awareness ability for multiple intents in the SIA-FTP framework can be attributed to the multi-label classification design of the IID model, enabling the model to extract comprehensive and informative features from ATC instructions.  

{{In summary, the visualization results demonstrate that the proposed SIA-FTP learns expected flight transition patterns in instruction-driven maneuvering flight phases, which also supports our motivations for explicitly integrating the spoken ATC instructions into the FTP procedures.}}

\subsection{Ablation Study} \label{sec3.8}
To further validate the advantages of the 3-stage progressive multi-modal learning paradigm in the proposed SIA-FTP framework, two further ablation studies are designed in this section as follows. 
\begin{itemize}
    \item \textbf{A1}: To validate the effectiveness of the IID finetuning (Stage 2-2), in this experiment, we directly utilize the output of the BERT pre-training after Stage 2-1 as the sentence-level intent embeddings $\mathbf{SI}_{emb}$, while other setting remains unchanged with original SIA-FTP framework.   
    In other words, instead of the finetuning with the IID model in Stage 2-2, only Stage 2-1 is conducted to learn the instruction embeddings in the intent-oriented instruction embedding learning phase. 

    \item \textbf{A2}: To validate the effectiveness of the intent-oriented instruction embedding learning, Stage 2 is fully removed to develop the SIA-FTP in this experiment. 
    Specifically, the BERT model is employed as the instruction encoder to build the multi-modal SIA-FTP model after the trajectory-based FTP pre-training and 
    proceeds directly to Stage 3 for model training with paired trajectory-instruction data. 
\end{itemize}

\begin{table}[htbp]
    \centering
    \setlength\tabcolsep{4 pt} 
    \caption{{{The experimental results of the ablation study. Notation w/o indicates the specific module is not included in the SIA-FTP.}}} \label{tab_ab6}
    \begin{tabular}{c|c|c|ccc|ccc|ccc|c}
    \toprule
    \multirow{2}{*}{\textbf{Exp.}} &\multirow{2}{*}{\textbf{Methods}}                    & \multirow{2}{*}{\textbf{Horizon}} & \multicolumn{3}{c|}{\textbf{MAE} $\downarrow$} & \multicolumn{3}{c|}{\textbf{MAPE (\%)} $\downarrow$} & \multicolumn{3}{c|}{\textbf{RMSE} $\downarrow$} & \multirow{2}{*}{\textbf{MDE} $\downarrow$} \\ \cline{4-12}
                                   &                      &                                   & Lon       & Lat       & Alt      & Lon          & Lat         & Alt       & Lon        & Lat       & Alt      &                               \\ \midrule
    \multirow{4}{*}{\textbf{A1}}   &\multirow{4}{*}{\makecell{SIA-FTP\\(w/o\\Stage 2-2)}} & 1 & 0.0020    & 0.0016    & 0.79     & 0.0019       & 0.0060      & 0.10      & 0.0037     & 0.0030    & 2.35     & 0.17                          \\
                                   &                      & 3                                 & 0.0037    & 0.0031    & 4.14     & 0.0034       & 0.0114      & 0.46      & 0.0066     & 0.0057    & 7.46     & 0.32                          \\
                                   &                      & 9                                 & 0.0094    & 0.0085    & 8.34     & 0.0088       & 0.0317      & 0.92      & 0.0183     & 0.0167    & 15.50    & 0.83                          \\
                                   &                      & 15                                & 0.0149    & 0.0139    & 10.55    & 0.0139       & 0.0519      & 1.15      & 0.0298     & 0.0271    & 20.34    & 1.32                          \\ \hline
    \multirow{4}{*}{\textbf{A2}}   &\multirow{4}{*}{\makecell{SIA-FTP\\(w/o \\Stage 2)}} & 1  & 0.0018    & 0.0015    & 0.76     & 0.0017       & 0.0057      & 0.09      & 0.0032     & 0.0030    & 2.37     & 0.16                          \\
                                   &                      & 3                                 & 0.0034    & 0.0029    & 3.92     & 0.0031       & 0.0110      & 0.44      & 0.0058     & 0.0055    & 7.25     & 0.30                          \\
                                   &                      & 9                                 & 0.0087    & 0.0088    & 7.99     & 0.0081       & 0.0329      & 0.88      & 0.0161     & 0.0180    & 15.48    & 0.80                          \\
                                   &                      & 15                                & 0.0137    & 0.0142    & 9.90     & 0.0127       & 0.0528      & 1.08      & 0.0262     & 0.0278    & 21.36    & 1.27                          \\ \bottomrule
    \end{tabular}
    \begin{tablenotes}
        \item[1] $\downarrow$ represents minimization indicators. 
    \end{tablenotes}
\end{table}

The experimental results of the ablation studies are presented in Table \ref{tab_ab6}. 
It can be seen that, compared to the original SIA-FTP framework, the performance of the A1 and A2 are degraded, which validates the effectiveness of the proposed training strategies in Stage 2. 
Fortunately, thanks to the fusion of the intent embeddings, their performance still surpasses that of the baseline models, which supports the motivation of this work again.

To be specific, the A2 model achieves higher performance even without intent-oriented instruction embedding learning stage. 
{{The above experimental results further demonstrate that integrating the ATC instruction is a promising solution to improve the performance of FTP tasks in instruction-driven maneuvering flight scenarios by explicit consideration of the driving factors (controlling intents).}} 
In addition, it can also be observed from the experimental results (A2 v.s. A1) that directly integrating the instruction embedding into SIA-FTP without any pre-training (A2) outperforms that of using instruction embedding obtained from pre-training BERT (A1). 
It can be attributed to the inability to learn intent-specific embeddings from universal textual representations without any further fine-grained optimizations through the IID task (Stage 2-2).

{{Furthermore, we also consider the MDE variations over 15 prediction horizons on the validation set across training epochs, as shown in Supplementary Figure 2.}} 
The results indicate that the intent-oriented instruction embedding learning stage can effectively enhance the fusion process between trajectories and textual instructions. 
Consequently, the convergence of the proposed SIA-FTP framework significantly sped up during the early epochs, achieving superior performance compared to the A1 and A2 models, which also demonstrates the advantages of the proposed 3-stage progressive learning paradigm.

\subsection{Interpretability Study}
\begin{figure}[!h]
	\centering 
	\includegraphics[width=0.98 \textwidth]{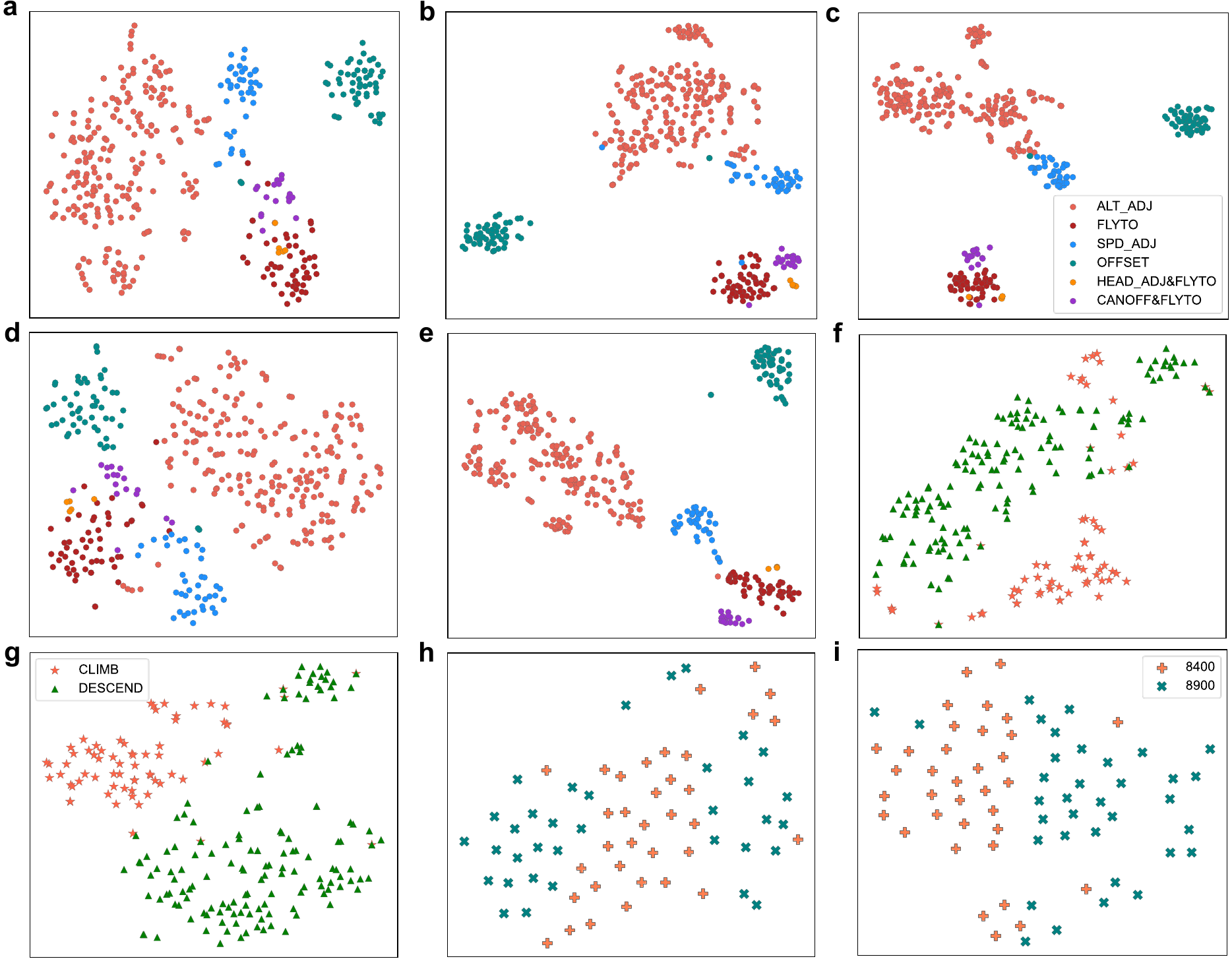}
	\caption{{Visualization of the text instruction embedding. \textbf{a-e} Visualization of the text instruction embedding with different maneuvering intents via the TSNE tool using different models. 
    a: BERT model (Stage 2-1). b: IID model (Stage 2-2). c: SIA-FTP model (Stage 1 to Stage 3). d: A1 model (SIA-FTP without stage 2-2). e: A2 model (SIA-FTP without stage 2). 
    \textbf{f-g} Visualization of the text instruction embedding with CLIMB and DESCEND intents via the TSNE tool using different models. f: IID model (Stage 2-2). g: SIA-FTP model (Stage 1 to Stage 3). 
    \textbf{h-i}Visualization of the text instruction embedding with different descending intents via the TSNE tool using different models. h: IID model (Stage 2-2). i: SIA-FTP model (Stage 1 to Stage 3). 
    Source data are provided as a Source Data file. }} \label{fig_vis}
\end{figure}

Analysis by the learned instruction embeddings. To further investigate the internal learning mechanisms in different stages, 
in this section, the t-Distributed Stochastic Neighbor Embedding (TSNE) tool is applied to cluster and visualize the sentence-level instruction embeddings $\mathbf{SI}_{emb}$ on the test set, which projects the learned high-dimension embeddings into 2D space. 
In general, a well-trained IID model should have the ability to distinguish the controlling intent of the ATC instructions clearly, 
i.e., the instructions with the same controlling intents should be clustered in the compact embedding space, as well as formulating a clear cluster gap with other intent embeddings.  
Specifically, the instruction embeddings $\mathbf{SI}_{emb}$ generated by the BERT-based pre-training model (Stage 2-1), IID model (Stage 2-2), SIA-FTP model (Stage 3), and the model of ablation experiments A1 and A2, are illustrated in Figure \ref{fig_vis}. 

{{As shown in Figure \ref{fig_vis}a, the BERT-based pre-trained model can only roughly distinguish general textual embeddings for different flight instructions and intents within the embedding space.}}
It can be attributed to the ATC instructions with the same controlling intents usually containing the same keywords and named entities, which inspires the pre-trained model to extract unified abstract embeddings. 
For instance, an ATC instruction with altitude adjustment intent must be conveyed the words "climbing" or "descending", as well as the flight level (e.g., eight thousand four hundred meters). 
With the finetuning of the IID task in Stage 2-1, the instruction embeddings within the same controlling intent are further compactly clustered in embedding space (Figure \ref{fig_vis}b). 
It is demonstrated that the intent-oriented instruction embedding finetuning stage promotes the model to extract the intent-related representations, not only for general text embeddings.  
Furthermore, as shown in Figure \ref{fig_vis}c, the embeddings of ATC instructions achieve high-compact cluster results in the embedding space after the multi-modal FTP finetuning of Stage 3,   
i.e., the distances of the samples within the same classes become closer, while the distances between classes are enlarged. 

It is also noted that, with the training stages progress, the samples with similar semantical intents, such as FLYTO and CANOFF\&FLYTO, or HEAD\_ADJ\&FLYTO, showcase increasingly closed embedding clusters in the feature space (Figure \ref{fig_vis}c). 
{{It also demonstrates that the model learns the semantic consistency (both kinematic and ATC instruction semantics) from the trajectory-instruction pairs during the multi-modal FTP finetuning, which further improved the capabilities of the informative intent embedding extraction.}}

In this section, the embeddings of the ATC instructions from the A1 and A2 models are also visualized in Figures \ref{fig_vis}d and Figures \ref{fig_vis}e, respectively. 
{{Compared to the original SIA-FTP model (Figures \ref{fig_vis}c), although the embeddings generated by the A1 and A2 models effectively distinguish samples in the embedding space among different intent classes, 
their embedding distributions lack compactness within each intent class.}} 
Therefore, it can be concluded that the above visualizations further validate the effectiveness and advantages of the proposed 3-stage progressive learning paradigm.

Insightful analysis. As observed from the visualization of trajectory prediction results in Section Visualization and Qualitative Analysis, 
the SIA-FTP model not only captures controlling intent but also attends to specific intent parameters from the ATC instructions, such as flight levels of the ALT\_ADJ intent and the miles of the OFFSET intent. 
{{To further examine the ability of the SIA-FTP model to distinguish specific instruction parameters within the same intent class, 
the instructions with ALT\_ADJ intent on the test set are further organized into two groups to perform the visualizations, i.e., climbing and descending.}}

The embeddings of the ALT\_ADJ instructions generated by the IID model and SIA-FTP model are visualized in Figures \ref{fig_vis}, respectively. 
As illustrated in Figures \ref{fig_vis}f-g, compared to the IID model, the SIA-FTP can further learn altitude-specific features between climbing and descending instructions through multi-modal FTP finetuning with trajectory-instruction pairs.

Similarly, to further examine the capability of the model to consider instruction parameters, the samples involving the common flight levels, including "descending 8400" and "descending 8900", within the descending instructions are visualized in Figure \ref{fig_vis}h-i. 
It can be observed that the IID model fails to distinguish the samples clearly with different instruction parameters within the same class in the embedding space, since the task objective of the IID model is only to identify the intent of instruction. 
{{However, after the multi-modal FTP finetuning of stage 3, the SIA-FTP is able to distinguish the instructions with different parameters (flight level) in the embedding space, which provides interpretability for the higher performance of the proposed SIA-FTP framework.}}

\begin{figure}[!h]
	\centering 
	\includegraphics[width=0.7 \textwidth]{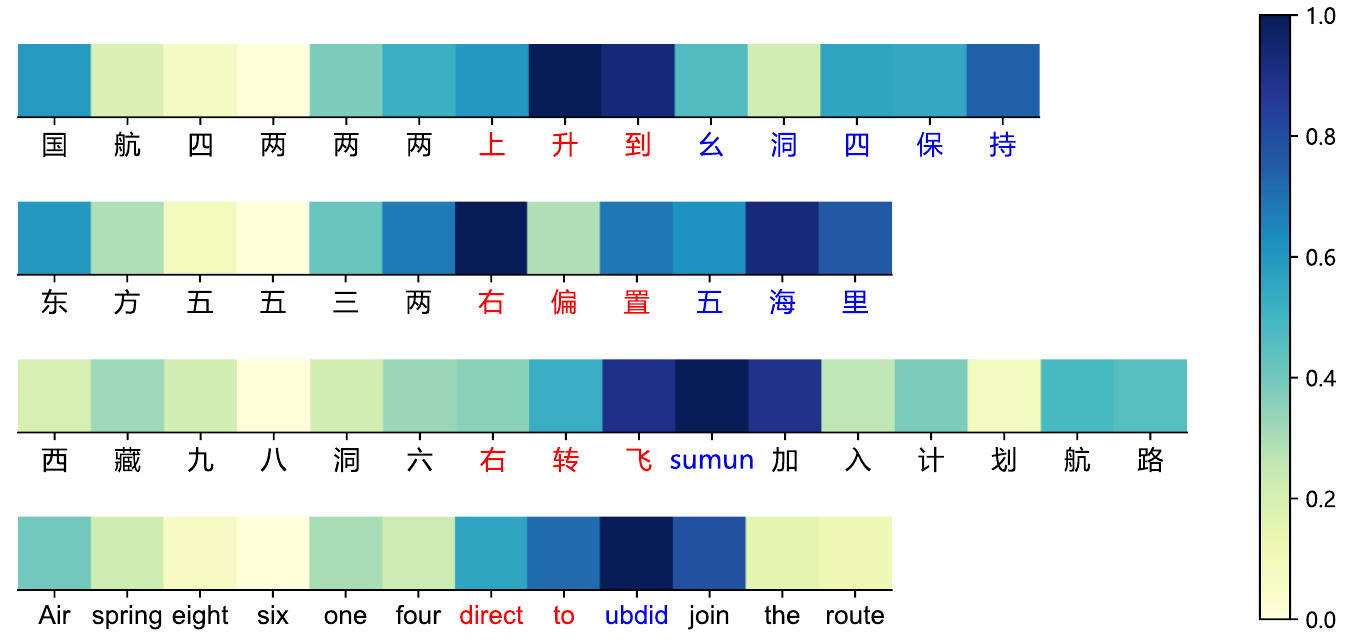}
	\caption{{The weights of the token in instruction embedding output by SIA-FTP framework. 
    The red fonts represent the keywords indicating intents, while the fonts with blue colors denote the critical parameters. 
    Source data are provided as a Source Data file.
    }} \label{fig_weight_vis}
\end{figure}

In this section, we also validate the attention weights of the tokens in ATC instructions during the multi-modal FTP prediction process. 
Figure \ref{fig_weight_vis} shows the token representations $\mathbf{H}_{SI}$ of 3 selected samples output by the SIA-FTP model. 
Specifically, $\mathbf{H}_{SI}$ is summed along the feature dimensions and quantifies the importance of tokens in ATC instruction.  
The darker colors represent larger attention weights, indicating the higher importance of tokens during the fusion process. 
It can be observed that the model not only focuses on intent-related keywords of the instructions but also assigns high weights to critical parameters to guide the model in predicting the flight trajectory toward target positional parameters (e.g., flight level, offset, waypoint, etc.).  

In summary, the above visualizations demonstrate that the proposed SIA-FTP has the ability to perceive the controlling intent and required parameters by integrating the textual ATC instructions. 
Moreover, the proposed 3-stage progressive learning paradigm is an effective way to fill the modality gap between the trajectory and textual instructions under the conditions of limited paired data.  
It is believed that the proposed SIA-FTP framework not only provides a solution for instruction-driven maneuvering ATC scenarios 
but also develops a powerful FTP tool to detect the potential risks of human factors in the ATC procedure.

\subsection{Generalization Study} 
\subsubsection{Integrating ATC instruction into LSTM+Attention} 
To validate the generalizability of integrating spoken instruction into the FTP procedure and the proposed 3-stage progressive learning paradigm, the LSTM+Attention model is applied to alternate the FlightBERT++ to conduct the SIA-FTP framework. 
Specifically, the LSTM+Attention baseline model is selected to perform the trajectory-based FTP pre-training in stage 1. 
Similar to stage 2 of the original SIA-FTP framework, the pre-trained IID model is used to extract intent-oriented instruction embedding. 
In the multi-modal finetuning stage, the instruction embedding is integrated into the LSTM+Attention model by the proposed multi-modal fusion mechanism during the step-by-step decoding process. 
The implementation details can be found in Supplementary Information (Section Implementation Details of the SIA-FTP Framework using LSTM+Attention).

\begin{figure}[!h]
	\centering 
	\includegraphics[width=\textwidth]{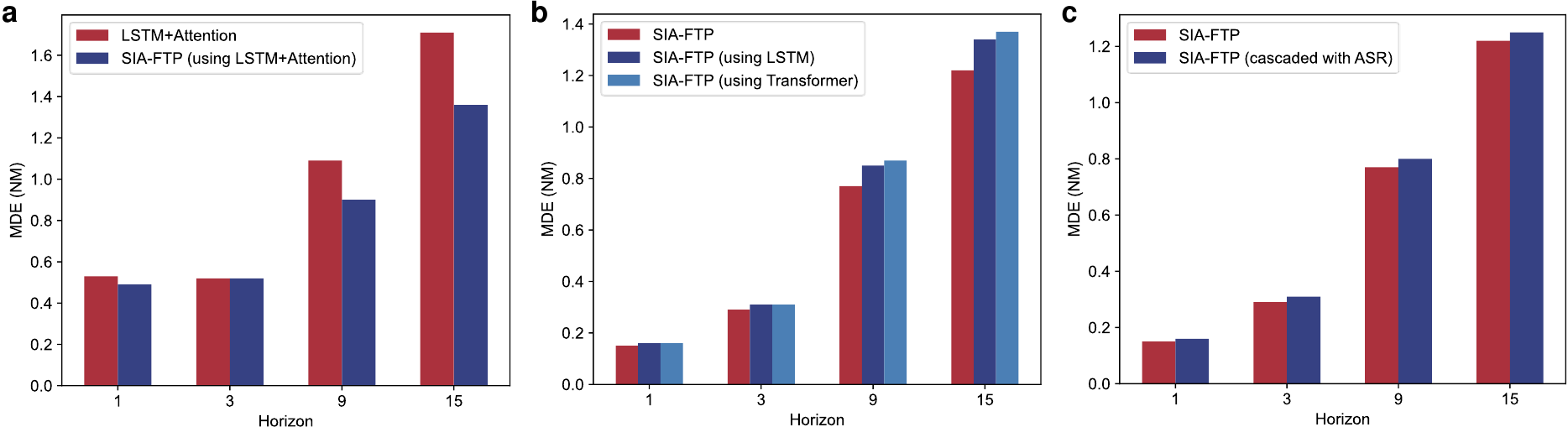}
	\caption{MDE comparison over the 1, 3,9 and 15 prediction horizons for generalization studies. \textbf{a} The MDE results of LSTM+Attention and SIA-FTP using LSTM+Attention in stage 1. 
    \textbf{b} The MDE results of the SIA-FTP, SIA-FTP using LSTM in stage 2, and SIA-FTP using Transformer in stage 2. 
    \textbf{c} The MDE results of the SIA-FTP and SIA-FTP cascaded with an ASR system. Source data are provided as a Source Data file.    
    } \label{fig_generalization}
\end{figure}

The experimental results are reported in Figure \ref{fig_generalization}a. 
It can be found that the LSTM+Attention model also achieves significant performance improvements by integrating the intent of spoken instruction using the proposed 3-stage training paradigm. 
Compared with the LSTM+Attention baseline, the SIA-FTP by LSTM+Attention achieves a 20.25\% relative MDE reduction in 15 prediction horizons (5 minutes). 
{{Moreover, by integrating the spoken instruction, the SIA-FTP by LSTM+Attention also obtained superior performance than FlightBERT++ in 9 (1.03 MDE) and 15 (1.55 MDE) prediction horizons, which further confirms the effectiveness of considering the controlling intent for multi-step FTP task.}} 
However, thanks to the powerful learning ability of the well-designed model architecture, the SIA-FTP by FlightBERT++ achieves higher prediction performance, which also validates our selection for the backbone network in this work. 
In summary, the experimental results of the generalization study not only demonstrate the effectiveness of integrating the intent of spoken instruction into FTP procedures and the 3-stage progressive learning paradigm, but also confirm the effectiveness and efficiency of the selected backbone network.

\subsubsection{Incorporating the ATC instruction into SIA-FTP using other language model architectures} 
Similarly, to further validate the effectiveness of the proposed SIA-FTP framework, the LSTM and Transformer architectures are also introduced to conduct the IID model in stage 2. 
Consequently, the pre-trained LSTM- or Transformer-based IID model is applied to alternate the BERT to build the multi-modal FTP model and extract the intent-oriented instruction embedded in stage 3. 
The implementation details and the experimental results of these IID models can be found in Supplementary Information (Evaluation of the Proposed IID Model). 

The comparison of MDE over 1, 3, 9, and 15 prediction horizons among the SIA-FTP using different language model architectures are shown in Figure \ref{fig_generalization}b. 
It is clear that the proposed SIA-FTP framework obtains expected performance improvement even when incorporating other types of language model architectures in stages 2 and 3. 
Furthermore, compared to the best baseline (FlightBERT++), the LSTM- and Transformer-based SIA-FTP framework achieve superior performance, obtaining about 13.5\% and 11.6\% MDE reduction over 15 prediction horizons, respectively. 
{{The results also confirm our primary motivation and contribution, i.e., integrating ATC instructions into the FTP model is a promising way to enhance prediction performance in instruction-driven maneuvering flight scenarios.}} 
Additionally, compared to the BERT model, it is observed that the SIA-FTP framework using LSTM and Transformer shows a slight performance reduction. 
This can be attributed to the better training strategy and network architecture of the BERT model, which facilitates learning the information ATC instruction embeddings.

\subsubsection{Evaluating the proposed SIA-FTP with ASR} 
As described in Section Introduction, in practice, the ASR serves as the front-end component of the SIA-FTP framework, which is employed to translate speech instructions into textual instruction. 
To validate the applicability of the proposed SIA-FTP framework in real-world scenarios, we cascade an ASR model with the SIA-FTP to evaluate the system-level performance. 
Specifically, an ASR method based on Wav2vec 2.0 from our previous work \cite{10106480} is employed as the front-end component, which is also trained on the M2ATS dataset. 
During the testing phase, instead of the manually annotated textual ATC instructions, the predicted text of the ASR model for the test set is used as the input for SIA-FTP. 

The ASR model achieves a 1.39\% character error rate (CER) on the test set, demonstrating considerable performance in the ATC domain. 
The FTP results are reported in Figure \ref{fig_generalization}c. 
It can be seen from the results that the SIA-FTP achieves comparable performance with only a slight reduction when cascaded with an ASR model in the front, compared to using golden annotations as input. 
It is observed that the ASR performance is critical to the overall effectiveness of the SIA-FTP framework, in which the misrecognition of intent-related keywords by the ASR can impact the semantics of the extracted intent-oriented instruction embeddings. 
Consequently, cascading SIA-FTP with a high-performance ASR is essential for ensuring reliable application in real-world scenarios. 
Fortunately, thanks to advancements in ASR techniques in the ATC domain, the proposed SIA-FTP framework has great potential for real-world ATC applications.

\section{Discussion} \label{sec4}
{{In this work, an instruction-driven FTP paradigm is proposed to incorporate spoken instructions into the ATC automation process, which provides a promising solution to detect the potential risks caused by human factors in real-time ATC operations. 
To this end, a SIA-FTP framework is proposed to consider the spoken instruction in the FTP procedure and implement a multi-modal FTP model within the controlling instruction-driven maneuvering flight phase. 
To address the modality gap between the textual spoken instructions and flight trajectory, we decompose the multi-modal learning of FTP tasks into 3 stages, including trajectory-based FTP pre-training, intent-oriented instruction embedding learning, and multi-modal finetuning. 
The joint model is optimized with the limited trajectory-instruction pairs to further learn the flight transition patterns under instruction-driven maneuvering flight scenarios.}} 
The effectiveness of the proposed framework is demonstrated by extensive experiments based on a real-world dataset, and all the proposed strategies contribute to desired performance improvements. 
The multi-horizon prediction task is achieved with considerable performance. 
Most importantly, the intent and required position-guided parameters can be accurately perceived to enhance the FTP task. 
The fusion process between trajectory and text instruction is intuitively understood by extensive visualization results.

Although the SIA-FTP framework demonstrates significant performance improvements over comparative baselines, Figure \ref{fig_vis}c shows that a few samples are difficult for the SIA-FTP framework to distinguish in the embedding space. 
In practice, the intent misclassification of ATC instructions might potentially impact prediction errors of the proposed SIA-FTP framework. 
By in-depth analysis of the samples in the dataset, it is found that the challenging or misclassified samples in Figure \ref{fig_vis}c can be categorized into two primary classes: 

\begin{enumerate}
    \item Figure \ref{fig_vis}c shows that some instruction embeddings with SPD\_ADJ intent are in close proximity to those with ALT\_ADJ intents in feature space. 
    This can be attributed to SPD\_ADJ intents often being issued during the execution of ALT\_ADJ intents. 
    In the $3^{rd}$ stage of the proposed SIA-FTP framework, the multi-modal FTP model learns both kinematic and controlling intent semantics from real-time observed trajectories and ATC instructions. 
    Consequently, the intent-oriented instruction embeddings are fine-tuned by kinematic semantics during stage three, resulting in the proximity of ALT\_ADJ and SPD\_ADJ instruction embeddings in feature space. 
    {{This confirms that the proposed 3-stage SIA-FTP framework can capture informative features from both real-time trajectory observations and ATC instructions, thereby enhancing FTP performance in instruction-driven maneuvering flight processes.}} 
    Additionally, Supplementary Figure 3a visualizes a representative sample where an ATCo issues a speed adjustment instruction during the climbing phase. 
    In this case, the instruction embedding extracted by the SIA-FTP framework is clustered close to the ALT\_ADJ intents in feature space, but the SIA-FTP still achieves the expected prediction results due to the learned informative features.
    
    \item The ATC instructions with multiple intents are not distinctly separated from single-intent instructions in Figure \ref{fig_vis}c. 
    Analysis shows that when multiple intents are presented in ATC instructions, the SIA-FTP framework tends to focus on the "primary" intent. 
    For instance, the instructions contain CANOFF\&FLYTO (cancel offset and direct to) and HEAD\_ADJ\&FLYTO (heading adjustment and direct to) intents, the SIA-FTP framework typically pays more attention to the FLYTO intent since it is the main goal of the ATC instruction, 
    with CANOFF and HEAD\_ADJ being accompanying actions during the execution of FLYTO. 
    This observation also can be confirmed by the weight of the token in $3^{rd}$ samples of Figure \ref{fig_weight_vis}, where the token weights of the FLYTO intent ("direct to sumun") are higher than the HEAD\_ADJ intent ("turn right"). 
    Supplementary Figure 3b visualizes a representative sample with multiple intents that are not correctly classified in Figure 4c to investigate the prediction errors of the SIA-FTP framework. 
    It shows that the SIA-FTP mainly focuses on the FLYTO intent with marginal considerations on the CANOFF intent (cancel offset). 
    Although the SIA-FTP model provides delayed prediction to timely cancel the offset, it still achieves comparable prediction performance (1.46 MDE). 
    {{Most importantly, the predicted trajectory also approaches the waypoint "sumun" by considering the "primary" FLYTO intent, which is the prominent advantage (instruction-driven) compared to the conventional data-driven FTP model.}}
\end{enumerate}

In summary, we present a perspective to consider human factors in real-time ATC operations by developing an instruction-driven FTP model and propose a solution to fuse the spatial-temporal trajectory data and textual instructions under limited data scale. 
It is believed that the proposed SIA-FTP framework can not only empower modern ATC systems to enhance aviation safety but also provide technical insights for multi-modal ATC data processing.

In the future, we will focus on exploring an efficient fusion of spoken instruction and flight trajectory. 
In addition, extracting more informative and detailed maneuvering intents from the spoken instructions is also an interesting research topic.

\section{Methods} 
\subsection{Overview of the proposed framework} 
\begin{figure}[]
	\centering
	\includegraphics[width=0.98\textwidth]{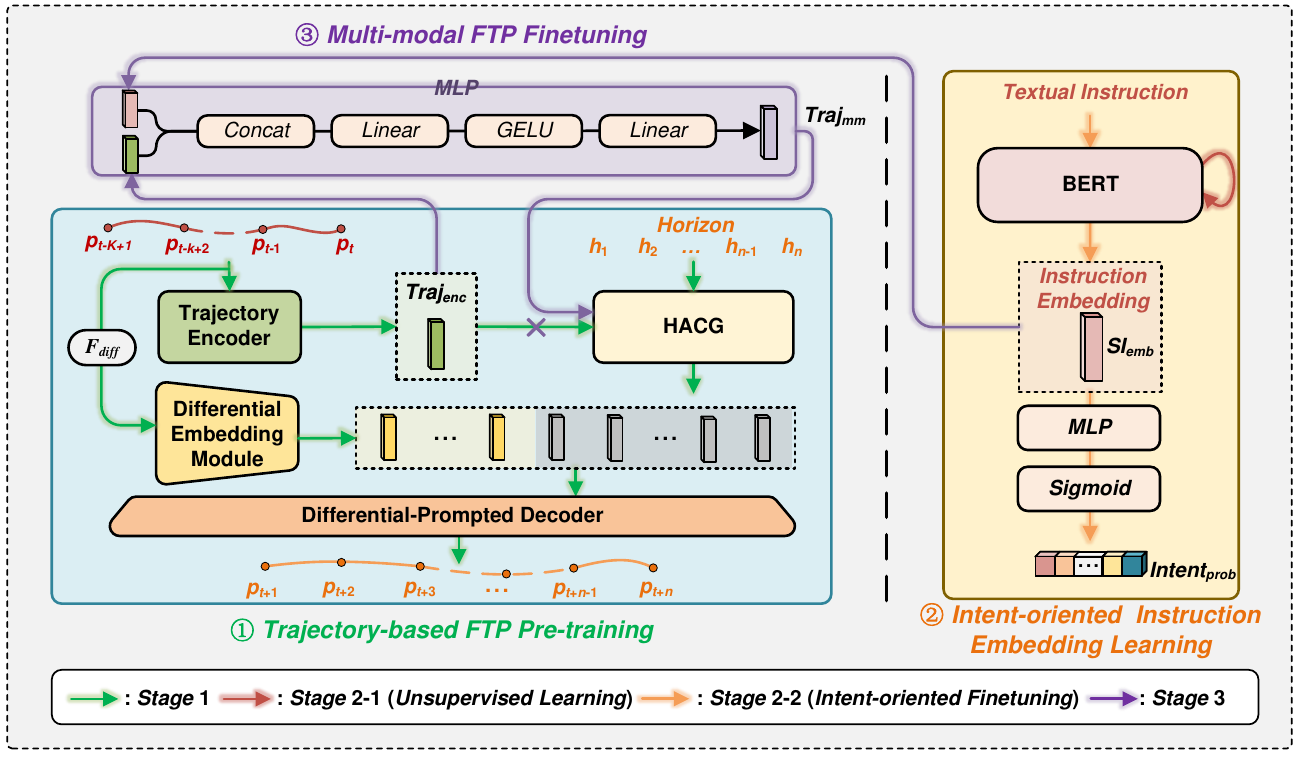}
	\caption{Overview of the proposed SIA-FTP framework. The $F_{diff}$ represents the differential operator that is applied to obtain the differential sequence of the observation. 
    HACG: horizon-aware context generator. Concat: concatenation operation. GELU: activation function. MLP: multi-layer perception.}  
	\label{fig1}
\end{figure}
The architecture of the proposed SIA-FTP framework is illustrated in Figure \ref{fig1}. 
Specifically, the proposed framework decomposes the learning procedure of the FTP tasks into three stages to address the modality gap and low-resource problem of trajectory-instruction pairs, 
including trajectory-based FTP pre-training, intent-oriented instruction embedding learning, and multi-modal finetuning. 

In the trajectory-based FTP pre-training stage, the FlightBERT++ is employed to train a data-driven FTP model based on motion patterns of the trajectory sequence (without considering controlling intents), which aims to learn the typical spatial-temporal flight transition patterns by leveraging historical flight trajectory data.
In this stage, the robust spatial representations of trajectory points and generalized flight motion patterns are expected to be learned through the training process, 
thereby establishing a robust FTP foundation model and ensuring effective generalization for common flight scenarios.
{{Although a generalized FTP model can be obtained in this stage, it may not be robust enough to handle instruction-driven maneuvering flight scenarios.}} 

For the intent-oriented instruction embedding learning stage, a multi-label intent identification (IID) model is proposed to learn the discriminative features of spoken instruction within the different controlling intents. 
{{To this end, a BERT-based neural architecture is designed to perform intent identification, whose effectiveness has been demonstrated in many Natural Language Processing (NLP) tasks. 
The primary purpose of this stage is to project the textual ATC instructions into a shared embedding space (stage 2-1) and extract informative instruction embeddings that capture the strong semantics of the controlling intents (stage 2-2), thereby laying a foundation for bridging the modality gap between textual instructions and trajectory data.
}}

In the multi-modal FTP finetuning stage, a multi-modal fusion mechanism is designed to incorporate the instruction embeddings into the pre-trained FTP models and to combine the pre-trained FTP and IID network into a joint model. 
This mechanism conducts a bridge of the data flow for the spoken instruction and trajectory modal, which enables us to finetune the joint model by utilizing the trajectory-instruction pairs. 
{{By combining information from both modalities, this stage ensures the multi-modal FTP model can learn both kinematic and controlling intent semantics from real-time observed trajectories and ATC instructions, enabling FTP for instruction-driven high-maneuvering flight scenarios.}} 

By decomposing the learning procedures, the model parameters are mainly updated in the separate training stage and jointly optimized in the multi-modal finetuning stage. 
On the one hand, we can fully leverage the well-resourced unimodal data to learn flight transition patterns and instruction embeddings separately. 
On the other hand, the trajectory and spoken instruction with different modalities are projected into the latent feature space in the $1^{st}$ and $2^{nd}$ stages, and further fused into a unified feature space in the $3^{rd}$ stage. 
Compared to fusing the multi-modal data directly, it is believed that the proposed model and training strategy can easily learn the consistent semantics between trajectories and spoken instructions in the latent feature space. 
{{In this way, it is expected to develop a multi-modal FTP model that is robust to instruction-driven high-maneuvering scenarios even with limited trajectory-instruction pairs.}} 
A detailed description of each stage is provided in the following sections. 

\subsection{Trajectory-based FTP pre-training} \label{sec5.2}
In general, once the ATC instruction reaches a consensus between ATCo and aircrews, the pilot will proceed to maneuver the aircraft following the controlling intent of instruction. 
The maneuvering process requires a sustained execution from the initial operation to the completion, typically an average of 1-5 minutes. 
{{Therefore, in this work, the instruction-driven FTP is a multi-horizon forecasting task to predict the flight trajectory of the entire instruction execution process.}} 
In this context, the SIA-FTP task requires the FTP model can fit the complicated patterns in a data-driven manner, for both the trajectory evolution patterns and the resulting maneuvering operations of the controlling instructions.
{{Considering that FlightBERT++ \cite{guo2023flightbert++} achieves superior performance on both precision and efficiency within the multi-horizon FTP approaches, the FlightBERT++ serves as the prediction framework to conduct the instruction-driven FTP task.}}
In this section, a brief review of FlightBERT++ is provided to elaborate on the model architectures. 

Specifically, the FlightBERT++ is composed of three modules, i.e., trajectory encoder (TE), horizon-aware context generator (HACG), and differential-prompted decoder (DPD). 
The goal of the TE module is to project the observation trajectory into a trajectory-level high-dimensional representation. 
In succession, the HACG module is designed to generate multi-horizon context embeddings based on the high-dimensional representation and the prior horizon information $\mathrm{horizon} = \{h_1, h_2, ..., h_{n-1}, h_n\}$. 
These context embeddings and differential embeddings generated by the Conv1D-based differential embedding module (DEM) are jointly fed into the DPD module to predict the differential sequence of the future time steps. 
Finally, the Binary Cross Entropy (BCE) loss function is applied to optimize the FTP model in the training process. 

The FlightBERT++ inherits the powerful representation capacity of the binary encoding (BE) representation proposed in \cite{9945661}, 
and innovatively designed a series of technical improvements to solve the limitations of the BE representation, including high-bit prediction errors, error accumulation, and inefficient computation. 
The effectiveness of FlightBERT++ has been demonstrated in extensive diagnostic experiments, wherein even without external information, it can achieve high-confidence predictions in certain maneuvering scenarios. 
{{Therefore, we employ the FlightBERT++ architecture to conduct trajectory-based FTP pre-training and expect to achieve more precision and fine-grained prediction by incorporating spoken instruction.}} 
More details of the FlightBERT++ can be found in \cite{guo2023flightbert++}.

\subsection{Intent-oriented instruction embedding learning} \label{sec5.3}
As described in Section Introduction, the most critical driving factor of the flight status transitions in the ATC procedures is the spoken instruction with maneuvering intents. 
An intuitive idea is to extract the informative features that indicate the controlling intent from maneuvering instructions and integrate them into the FTP model, thereby improving the prediction precision. 

To this end, the spoken instructions are manually divided into 16 categories according to the requirements of ATC work. 
Furthermore, 6 maneuvering instructions that significantly impact flight status are utilized to conduct the SIA-FTP framework, as listed in Table \ref{tab1}. 
Obviously, it is essential to first identify the maneuvering instruction, thereby extracting the discriminative features to support the FTP tasks. 
In real-world ATC works, an instruction usually with over 1 controlling intent, brings a challenge to capture the discriminative intent features. 
For instance, the instruction of No. 6 in Table \ref{tab1} indicates the left flight turn and direct to the ubdid (waypoint). 

\begin{table}[h]
    \caption[]{The descriptions of selected 6 categories maneuvering instructions} \label{tab1}
    \begin{tabular}{ccll}
    \toprule
    \textbf{No.} & \textbf{Intents} & \multicolumn{1}{c}{\textbf{Descriptions}}            & \multicolumn{1}{c}{\textbf{Example}}                                                                                           \\ \midrule
    1            & ALT\_ADJ         & The instruction indicates the altitude changes.    & \begin{tabular}[c]{@{}l@{}}Air China nine three one climb maintain \\ one two thousand two hundred meters\end{tabular}   \\
    2            & OFFSET           & The instruction indicates the offset of the track. & \begin{tabular}[c]{@{}l@{}}Qatari eight four seven four offset three  \\  miles right of the track due to weather\end{tabular} \\
    3            & CANOFF           & The instruction indicates cancel offset.             & Loulan two six one six can cancel offset                                                                                       \\
    4            & FLYTO            & The instruction indicates direct to the waypoint.  & \begin{tabular}[c]{@{}l@{}}Dynasty five nine nine seven direct to  \\  marso\end{tabular}                                      \\
    5            & SPD\_ADJ         & The instruction indicates speed changes.             & \begin{tabular}[c]{@{}l@{}}West China six two five tree increase  \\  speed mach point seven six\end{tabular}                  \\
    6            & \begin{tabular}[c]{@{}l@{}}HEAD\_ADJ\\\&FLYTO\end{tabular} & The instruction indicates heading changes.         & \begin{tabular}[c]{@{}l@{}}Sichuan eight seven four five turn left  \\  direct to ubdid\end{tabular}  \\ \bottomrule
    \end{tabular}
\end{table}

Considering the aforementioned specificities of spoken instructions, a multi-label intent identification (IID) model is designed to learn the discriminative features. 
{{In this work, the IID model is also trained with two stages: i) (stage 2-1) unsupervised text representation learning and ii) (stage 2-2) intent-oriented instruction embedding finetuning.}}
Specifically, a BERT \cite{devlin2019bert} architecture is applied to build the IID model. 
In the first stage of the IID model (stage 2-1), the BERT-based network is trained to learn universal representations using a "masked language model" (MLM) pre-training objective by unlabeled text instructions. 
Let a textual instruction be $\mathrm{SI}=\{w_1, w_2, ..., w_l\}$, the BERT model can be formulated as Eq. (\ref{eq4}):
\begin{equation} \label{eq4}
    \mathbf{H}_\mathrm{SI} = \{h^{w_1}, h^{w_2}, ..., h^{w_l}\} = \operatorname{BERT}(\mathrm{SI}) 
\end{equation}
\noindent where $w_i, i=\{1, 2, ..., l\}$ is the $i^{th}$ basic token of the IID model, i.e., the Chinese characters for instructions using Mandarin and English words for English instructions. 
$l$ is the length of the instruction count by tokens. $h^{w_i} \in \mathbb{R}^{D}$ represents the high-dimensional features of $w_i$, and $D$ denotes the feature dimension.

In the stage of intent-oriented instruction embedding finetuning (stage 2-2), a multi-binary classification head is cascaded to the pre-trained BERT network to conduct the IID task. 
To capture the sentence-level instruction embeddings and retain more informative details, the token-level features are summed along the sequence dimensional to generate the final instruction embedding, as illustrated in Eq. (\ref{eq5}):  
\begin{equation} \label{eq5}
    \mathbf{SI}_{emb} = \operatorname{SUM}(\mathbf{H}_\mathrm{SI})
\end{equation}
\noindent where $\mathbf{SI}_{emb} \in \mathbb{R}^{D}$ server as the final instruction embedding, and the $SUM(\cdot)$ represents the sum operation. 

In succession, as illustrated in Eq. (\ref{eq6}), the instruction embedding $\mathbf{SI}_{emb}$ is fed to the multi-layer perception (MLP) module to generate the probabilities of each intent class $\mathbf{Intent}_{prob}$. 
Notably, the output of the MLP module is activated by the Sigmoid function due to the multi-label classification settings of the IID model. 
\begin{equation} \label{eq6}
    \mathbf{Intent}_{prob} = \operatorname{Sigmoid}(\operatorname{MLP}(\mathbf{SI}_{emb})) 
\end{equation}

The IID model is finetuned on the textual spoken instruction dataset with intent labels to learn the discriminative features for instructions with different intents. 
After the training process, the classification head is removed and the finetuned BERT model is preserved to obtain the sentence-level instruction embeddings $\mathbf{SI}_{emb}$.
It is believed that we can not only identify the maneuvering instructions but also extract the discriminative features in the instruction embeddings. 
{{In this way, the instruction embeddings can be further incorporated into FTP models to support instruction-driven FTP tasks.}} 

\subsection{Multi-modal FTP finetuning} \label{sec5.4}
In this stage, we mainly focus on efficiently incorporating instruction embeddings into the pre-trained FlightBERT++ model, allowing the FTP model to be aware of maneuvering controlling intent. 
As described in Section Introduction, one of the primary challenges of this work is without large-scale trajectory-instruction pairs to support multi-modal learning. 
Therefore, the new parameters of the modal fusion procedures should be as few as possible to reduce the difficulty of the finetuning process. 

To this end, firstly, the trajectory embedding output by the trajectory encoder is selected to fuse instruction embedding to generate the intent-aware trajectory embeddings, 
because the trajectory embedding implies the trajectory-level observation representation. 
Secondly, a simple yet effective multi-modal fusion mechanism is designed to incorporate the instruction embedding into FTP models and bridge the pre-trained FTP and BERT models. 

Specifically, as shown in Eq. (\ref{eq7}), a concatenate operation is performed to generate the multi-modal joint vector $\mathbf{J}_{mm} \in \mathbb{R}^M$ to roughly fuse trajectory embedding $\mathbf{Traj}_{enc}$ and instruction embedding $\mathbf{SI}_{emb}$, 
in which the $Concat[\cdot]$ is the function of concatenation, $M$ is the dimension of the $\mathbf{J}_{mm}$. 
\begin{equation} \label{eq7}
    \mathbf{J}_{mm} = \operatorname{Concat}[\mathbf{Traj}_{enc}, \mathbf{SI}_{emb}]
\end{equation}

Moreover, an MLP module is applied to conduct deep fusion of the two embeddings and project the fused vectors into original dimensions to generate the intent-aware trajectory embedding $\mathbf{Traj}_{mm}$, as Eq. (\ref{eq8}). 
\begin{equation} \label{eq8}
    \mathbf{Traj}_{mm} = \operatorname{MLP}(\mathbf{J}_{mm})
\end{equation}

Notably, instead of the $\mathbf{Traj}_{enc}$ in the trajectory-based FTP pre-training stage, the intent-aware trajectory embedding $\mathbf{Traj}_{mm}$ is used to generate the context embeddings by the HACG module. 
{{Finally, similar to the original FlightBERT++, the context embeddings are further fed into the DPD to generate predictions.}} 

Based on the above designs, we combine the pre-trained FTP model and BERT models into a joint model, only utilizing an MLP module with a few parameters. 
In the multi-modal FTP finetuning process, the joint model is optimized by the BCE loss function with the trajectory-instruction pairs. 
In this way, the inner correlation between flight transitions and instruction with maneuvering intents is expected to be learned, thereby enhancing the prediction ability of the FTP model in high-maneuvering scenarios. 

\section*{Data Availability}
We are not authorized to publicly release the whole dataset used during the current study concerning safety-critical issues. 
Nonetheless, the processed example samples are available at https://zenodo.org/records/13939556 \cite{gdy_scu_2024_13939556}. 
Source data are provided with this paper.

\section*{Code Availability}
The code is publicly available at https://zenodo.org/records/13939556 \cite{gdy_scu_2024_13939556}. 

\bibliographystyle{elsarticle-num}

\section*{Acknowledgements}
This work was supported by the National Natural Science Foundation of China (NSFC) under grants No. 62401380 (D.G. received this fund), 62371323 (Y.L. received this fund), U2333209 (Y.L. received this fund), 
and U20A20161 (H.Y., B.Y., J.Z., and Y.L. received this fund). 

\section*{Author Contributions Statement} 
D.G., Z.Z., B.Y., and Y.L. conceived and led the research project. 
D.G. and J.Z. developed the framework. 
D.G. and Y.L. devised neural architecture and wrote the paper. 
D.G. implemented the neural architecture and produced experimental results. 
D.G. and Z.Z. conducted data preprocessing and collected the experimental results. 
D.G., Y.L. and H.Y. wrote and edited the manuscript. 
All authors provided results discussions. 
Y.L., B.Y., and J.Z. approved the submission and accepted responsibility for the overall integrity of the paper.

\section*{Competing Interests Statement}
Authors declare no competing interests. 



\newpage
\begin{center}
   \LARGE  \textbf{Supplementary Information}
\end{center}


\setcounter{page}{1} 

\setcounter{section}{0}
\setcounter{equation}{0}
\setcounter{figure}{0}
\setcounter{table}{0}

\renewcommand{\figurename}{Supplementary Figure}
\renewcommand{\tablename}{Supplementary Table}

\section{Data Preprocessing}
The required trajectory attributes of the FTP task are extracted from the raw data to conduct the trajectory subset, including timestamp, callsign, longitude, latitude, altitude (LLA), and the velocity corresponding to the LLA dimensions. 
The preprocessing approaches of the trajectory data refer to our previous works, more details can be found in \cite{9945661, 3613759, guo2023flightbert++}. 
After the data preprocessing, a total of 10,362 trajectories were obtained in the subset with the region of the interest (ROI) $[94.616\degree, 113.689\degree]$, $[19.305\degree, 37.275\degree]$, $[0, 12500]$ for LLA, respectively. 
In this work, the trajectory subset is applied to the trajectory-based FTP pre-training stage (\textbf{Stage 1}) to train the FlightBERT++ model, as well as to train other FTP baselines. 

For the text instruction subset (\textbf{Stage 2}), the speech signal of the ATC communication procedure is segmented into discrete spoken instructions. 
Subsequently, the transcriptions, callsign, and controlling intents of each spoken instruction are manually labeled by expert annotators. 
Additionally, an amount of unlabeled text instructions is applied to pre-train the BERT model for the unsupervised text representation learning \cite{abs211102041}. 
After preprocessing and annotation, there are 113,862 text instructions in this subset to train the IID model in the intent-oriented instruction embedding learning stage. 

The trajectory-instruction subset (\textbf{Stage 3}) is conducted based on the above two subsets as the following process. 
Specifically, the text instructions with the 6 maneuvering intents are extracted from the text instruction subset, 
including ALT\_ADJ (altitude adjustment), OFFSET, CANOFF (cancel offset), FLYTO (direct to the waypoint), SPD\_ADJ (speech adjustment), HEAD\_ADJ (heading adjustment). 
The detailed description of the above maneuvering instructions is described in Section 4.3 of the Main Text. 
In succession, the trajectory segments are extracted from the trajectory subset according to the callsigns and timestamps for each text instruction. 
Finally, a total of 7520 trajectory-instruction pairs are matched to support the multi-modal FTP finetuning process. 

{{According to the statistics of instruction-driven maneuvering flight processes from the dataset, it is observed that most ATC instructions were completed in 5 minutes.}} 
Based on this observation and the problem formulation in Section 2.1 of the Main Text, in the trajectory-instruction subset, each trajectory segment contains 24 trajectory points with 20-second update intervals, including 9 observed and 15 future trajectory points. 
In the multi-modal finetuning stage, the 9 observed trajectory points and the textual ATC instruction serve as the inputs, which are fed into the conducted multi-modal FTP model together to finetune the parameters of pre-trained FTP and IID models. 
In other words, the past 3-minute trajectory and text instruction are applied to predict the flight trajectory in the future 5 minutes in stage 3.

\section{Definition of Evaluation Metrics}

The detailed definition of Mean Absolute Error (MAE), Mean Absolute Percentage Error (MAPE), Root of Mean Square Error (RMSE) are shown in Eq. (\ref{eq12}) - Eq. (\ref{eq15}). 
The Mean Deviation Error (MDE) metric is defined as Eq. (\ref{eq14}), the LLA is transformed into the earth-centered and earth-fixed (ECEF) coordinate system from the WGS-84 coordinate systems, 
and the Euclidean distance is calculated between the predictions and ground truth to evaluate the FTP performance. 

\begin{equation} \label{eq12}
  \mathrm{MAE} = \frac{1}{n} \frac{1}{h} \sum_{i=1} ^ n \sum_{j=1} ^ h   { \lvert a_{ij} - a^{\prime}_{ij}  \rvert }
\end{equation}

\begin{equation} \label{eq13}
  \mathrm{MAPE} = 100 \% \times \frac{1}{n} \frac{1}{h}  \sum_{i=1} ^ n \sum_{j=1} ^ h { \lvert \frac{a_{ij} - a^{\prime}_{ij}}{a_{ij}}  \rvert } 
\end{equation}

\begin{equation} \label{eq15}
  \mathrm{RMSE} = \sqrt{\frac{1}{n} \frac{1}{h} \sum_{i=1}^{n} \sum_{j=1} ^ h \left(a_{ij} - a^{\prime}_{ij} \right)^{2}}
\end{equation}

\begin{equation} \label{eq14}
  \mathrm{MDE} = \frac{1}{n} \frac{1}{h} \sum_{i=1} ^ n \sum_{j=1} ^ h \Phi( p_{ij} - p^{\prime}_{ij} ) 
\end{equation}

{{$n$ is the number of samples in the test set, and $h$ is the prediction horizon. $a, a^{\prime}$ are the real-value (decimals) of trajectory attributes in the ground truth and prediction, respectively.}} 
$p, p^{\prime}$ are the transformed values of the ground truth and prediction in the ECEF coordinate system. 
{{$\Phi(\cdot)$ is the calculation function of Euclidean distance (measured in nautical miles (NM)) in the 3D airspace.}}
In this work, we report the performance of the LLA attributes because these are the primary attributes of the trajectory point in the 4D FTP task and can uniquely determine the location of the flight.

\section{Results of Error distribution, MDE Curve and Statistical Test}

\begin{figure*}[h] 
  \centering
  \includegraphics[width=0.95 \textwidth]{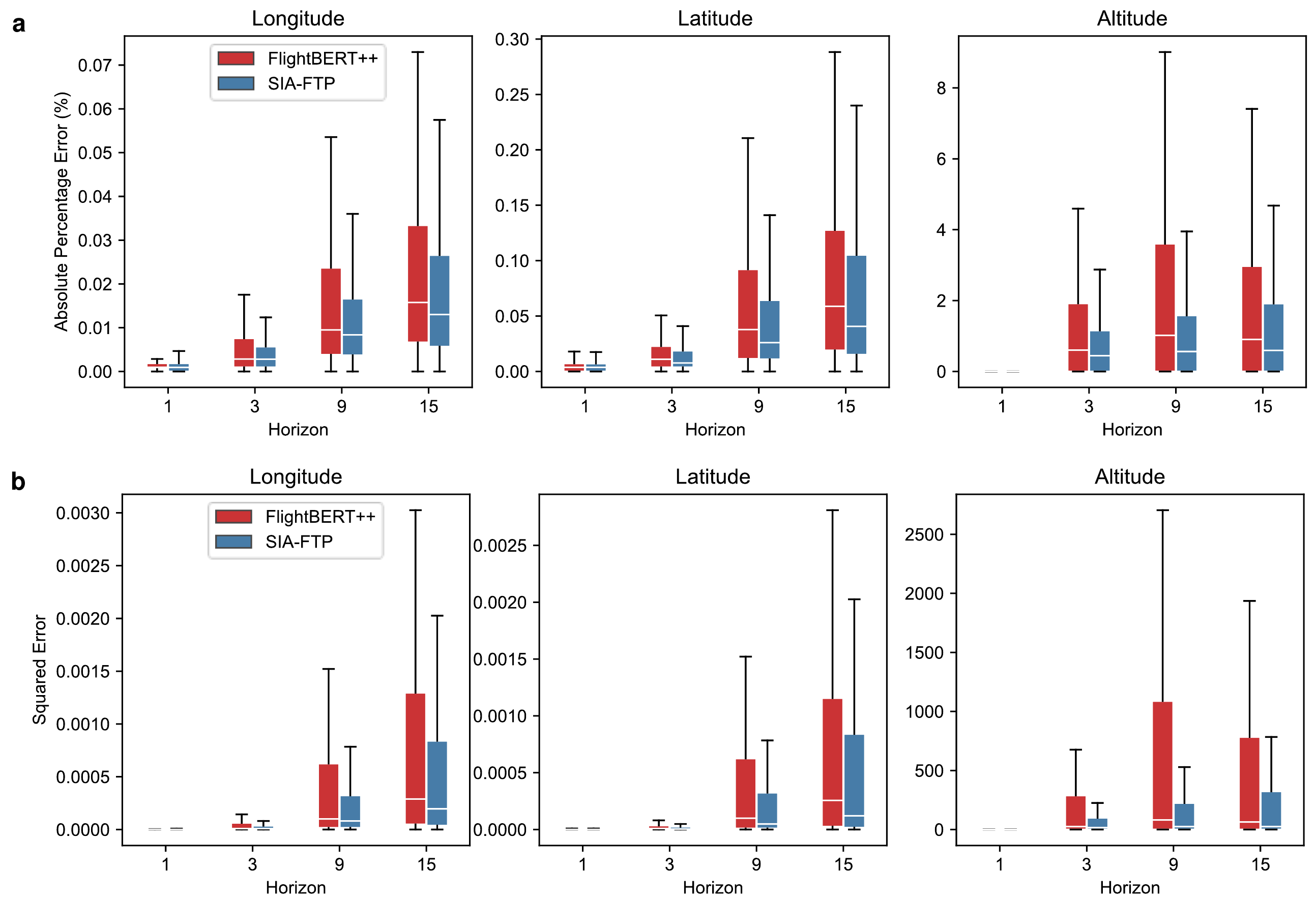}
  \caption{Error distribution of the FlightBERT++ and the proposed SIA-FTP. 
  \textbf{a} The distribution of absolute percentage error for LLA attributes in 1, 3, 9, and 15 prediction horizons. \textbf{b} The distribution of square error for LLA attributes in 1, 3, 9, and 15 prediction horizons.
  Boxplots of a and b show the median (center line), and 1st and 3rd quartiles (Q1 and Q3, respectively). 
  The error bars correspond to the Q1-(1.5*IQR) and Q3 + (1.5*IQR) range (IQR = Inter-Quartile Range). 
  Data points below Q1 – (1.5*IQR) or above Q3 + (1.5*IQR) are considered outliers and not shown in the boxplots.
  Source data are provided as a Source Data file.} \label{fig_err_dis}
\end{figure*} 

{{
To further validate the effectiveness of SIA-FTP, a statistical test is conducted to compare the significant differences between SIA-FTP and FlightBERT++. 
Table \ref{tab:pvalues} shows the p-values obtained from the Wilcoxon signed-rank test across different trajectory attributes (Longitude, Latitude, and Altitude) and 
four error metrics (Absolute Error, Absolute Percentage Error, Squared Error, and Deviation Error). 
The results indicate significant differences between the SIA-FTP and FlightBERT++ models in prediction horizons of $3^{rd}$, $9^{th}$, and $15^{th}$. 
Specifically, the p-values for Longitude, Latitude, and Altitude are consistently below the 0.05 significance threshold for these prediction horizons. 
It is demonstrated that the SIA-FTP consistently outperforms FlightBERT++ across longer horizons, with p-values approaching 0. 

\begin{table}[h]
  \caption{{{The p-values of each trajectory attribute across four metrics using the Wilcoxon signed-rank test.}}} \label{tab:pvalues}
  \begin{tabular}{c|ccc|ccc|ccc|c}
  \toprule
  \multirow{2}{*}{\textbf{Horizon}} & \multicolumn{3}{c|}{\textbf{Absolute Error}}   & \multicolumn{3}{c|}{\textbf{Absolute Percentage Error}} & \multicolumn{3}{c|}{\textbf{Square Error}}   & \multirow{2}{*}{\textbf{\makecell{Deviation\\Error}}} \\ \cline{2-10}
                                    & Lon    & Lat    & Alt    & Lon      & Lat     & Alt     & Lon    & Lat    & Alt    &                     \\ \midrule
  1                                 & 0.0525 & 0.7592 & 0.0583 & 0.0726   & 0.9395  & 0.0806  & 0.0535 & 0.8412 & 0.0534 & 0.2521              \\
  3                                 & 0.0003 & 0.0042 & 0.0000 & 0.0001   & 0.0039  & 0.0000  & 0.0004 & 0.0068 & 0.0000 & 0.0000              \\
  9                                 & 0.0000 & 0.0000 & 0.0000 & 0.0000   & 0.0000  & 0.0000  & 0.0000 & 0.0000 & 0.0000 & 0.0000              \\
  15                                & 0.0000 & 0.0003 & 0.0000 & 0.0000   & 0.0003  & 0.0000  & 0.0000 & 0.0004 & 0.0000 & 0.0000              \\ \bottomrule
  \end{tabular}
\end{table}

However, for the $1^{st}$ horizon, the p-values across all attributes are higher than 0.05, demonstrating that the difference between SIA-FTP and FlightBERT++ models is not statistically significant. 
As discussed in the Main Text, this result can be attributed to the limited influence of instruction-driven factors on flight trajectory in the first prediction horizon. 
Since FlightBERT++ serves as the trajectory representation learning component of SIA-FTP, it achieves comparable performance in the early horizons of prediction. 
Furthermore, analysis of the dataset reveals that in some cases, pilots may not have executed flight instructions within the $1^{st}$ prediction horizon (20 seconds) due to operational habits, further contributing to the observed similarity in performance.

In summary, the statistical analysis confirms that SIA-FTP significantly outperforms FlightBERT++ in predicting flight trajectories over longer horizons, while both models demonstrate comparable performance in the $1^{st}$ horizon. 
}}

\begin{figure}[h]
	\centering 
	\includegraphics[width=0.45 \textwidth]{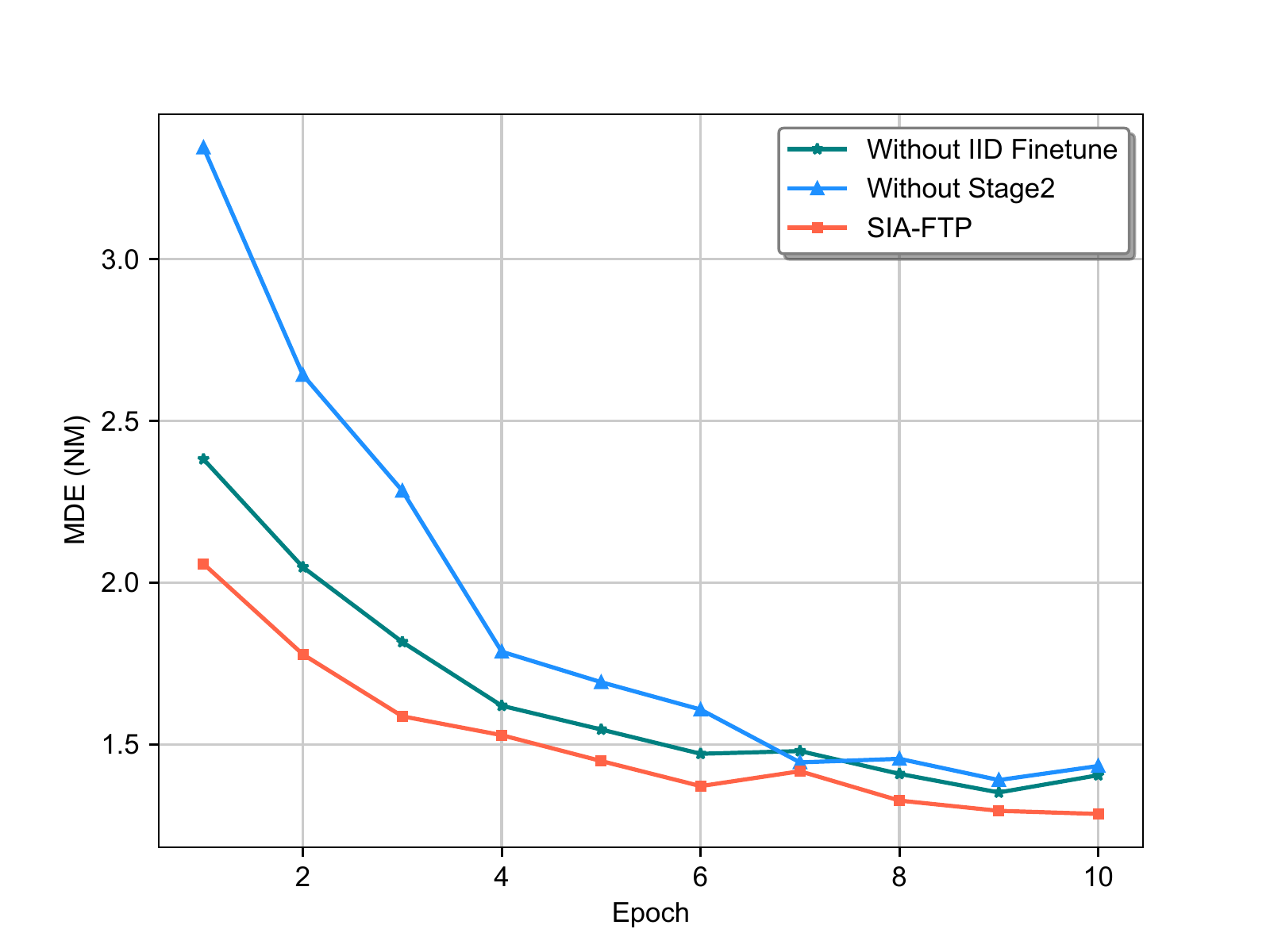}
	\caption{The MDE curve of the 15 prediction horizons for A1, A2 and SIA-FTP models. Source data are provided as a Source Data file.} \label{fig_mde6}
\end{figure}

\begin{figure*}[h]
	\centering
	\includegraphics[width=0.98\textwidth]{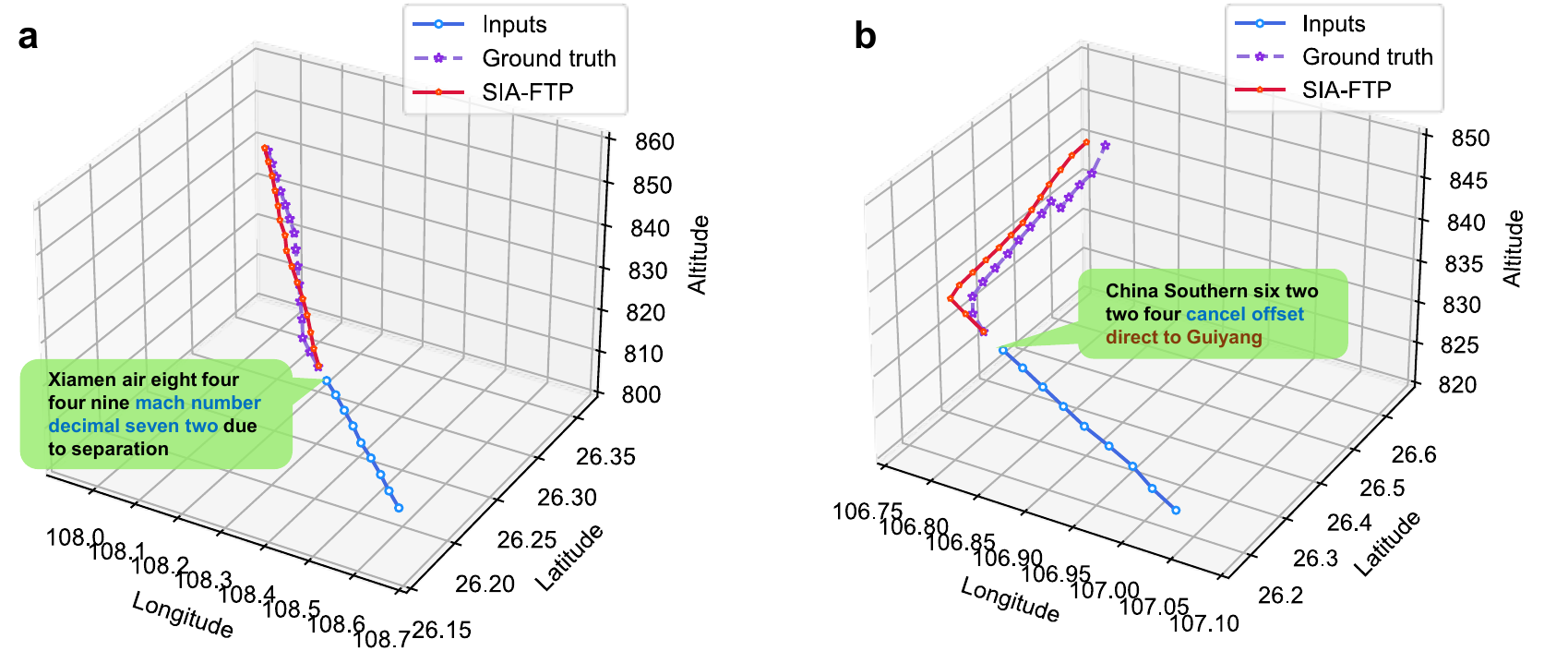}
	\caption{Case study of the samples that are not well clustered in Figure 5c. \textbf{a} The sample with SPD\_ADJ intent but it is close to ALT\_ADJ samples in feature space. \textbf{b} The sample with CANOFF\&FLYTO intents but clustered into FLYTO categories in feature space.
  Source data are provided as a Source Data file.}
	\label{fig:case}
\end{figure*}

\section{Evaluation of the Proposed IID Model}
To validate the effectiveness of the proposed IID model for the spoken instruction intent identification tasks, 
3 competitive baselines are selected to conduct the comparisons, as listed below.

\begin{itemize}
    \item \textbf{LSTM}: In this experiment, a word embedding layer is used to learn the high-dimensional features of the tokens. 
    In succession, a 2-layer LSTM network with 128 neurons is employed to capture the sequence correlations of the spoken instruction. 
    Similar to the proposed IID model, an FC layer with a Sigmoid activation function is applied to perform the multi-label classification.  
    \item \textbf{Transformer}: Compared to the aforementioned LSTM method, 4 Transformer blocks are cascaded to build the backbone of the IID model.  
    \item \textbf{BERT (Without pre-training)}: To validate the effectiveness of the unsupervised text representation learning (Stage 2-1) for the IID task, 
    in this experiment, the BERT architectures and training strategy are solely to construct the IID model and skip the unsupervised pre-training phase. 
\end{itemize}

\begin{table}[htbp]
    \centering
    \setlength\tabcolsep{20 pt} 
    \caption{The experimental results of the proposed IID model and baselines.} \label{tab6-iid-result}
    \begin{tabular}{ccccc} 
    \toprule
    \textbf{Methods}             & \textbf{Precision} \% & \textbf{Recall} \% & \textbf{F1 Score}  \% & \textbf{Accuracy} \% \\ \hline
    LSTM                         & 98.42              & 98.15           & 98.07             & 97.87        \\
    Transformer                  & 98.18              & 98.15           & 97.75             & 97.29        \\ 
    BERT                         & 99.68              & 99.45           & 99.54             & 99.20        \\
    IID (The proposed)           & \bf{99.76}         & \bf{99.70}      & \bf{99.72}        & \bf{99.50}   \\
    \bottomrule
    \end{tabular}
    \begin{tablenotes}
      \item[1] The bold ones denote the best performance on the corresponding metric.
  \end{tablenotes}
\end{table}

The experimental results are reported in Table \ref{tab6-iid-result}. 
A total of 4 common metrics used in multi-label classification tasks are employed to evaluate the model performance, including precision, recall, F1 score, and accuracy. 
The experimental results demonstrate that all the above IID models have achieved the desired performance, with the accuracy of intent identification exceeding 97\%. 
On the one hand, the ATC instructions subject to the communication rules recommended by the International Civil Aviation Organization (ICAO), 
and the controlling intents are generally characterized by specific keywords in the ATC instructions. 
For instance, the ATC instructions with ALT\_ADJ intents (altitude adjustment) usually contain terms like 'climb' or 'descend', while instructions with SPD\_ADJ intents (speed adjustment) are usually mentioned with 'speed' or 'Mach'. 
On the other hand, for ATC instructions without intent keywords (not strictly following the communication rules of the ICAO), 
the models are capable of inferring the controlling intents based on other elements in the instructions. 
For example, instructions with flight levels (such as 'eight thousand four', and 'eight thousand nine') are often associated with ALT\_ADJ intents, 
and those mentioned waypoints generally correspond to FLYTO intents. 

Moreover, it can be found from the results that the BERT model significantly outperforms the LSTM and Transformer models across the 4 metrics. 
Compared with LSTM and Transformer models, the BERT architecture has a larger number of parameters, 
and its Masked Language Model (MLM) training strategy is more conducive to learning the underlying correlations between tokens in the instructions. 
Benefiting from the unsupervised text pre-training, the proposed IID model achieved the best performance, reaching an accuracy of 99.50\%. 

\section{Implementation Details of the SIA-FTP Framework using LSTM+Attention} \label{lstmatt}

\begin{figure}[t]
	\centering
	\includegraphics[width=0.9\textwidth]{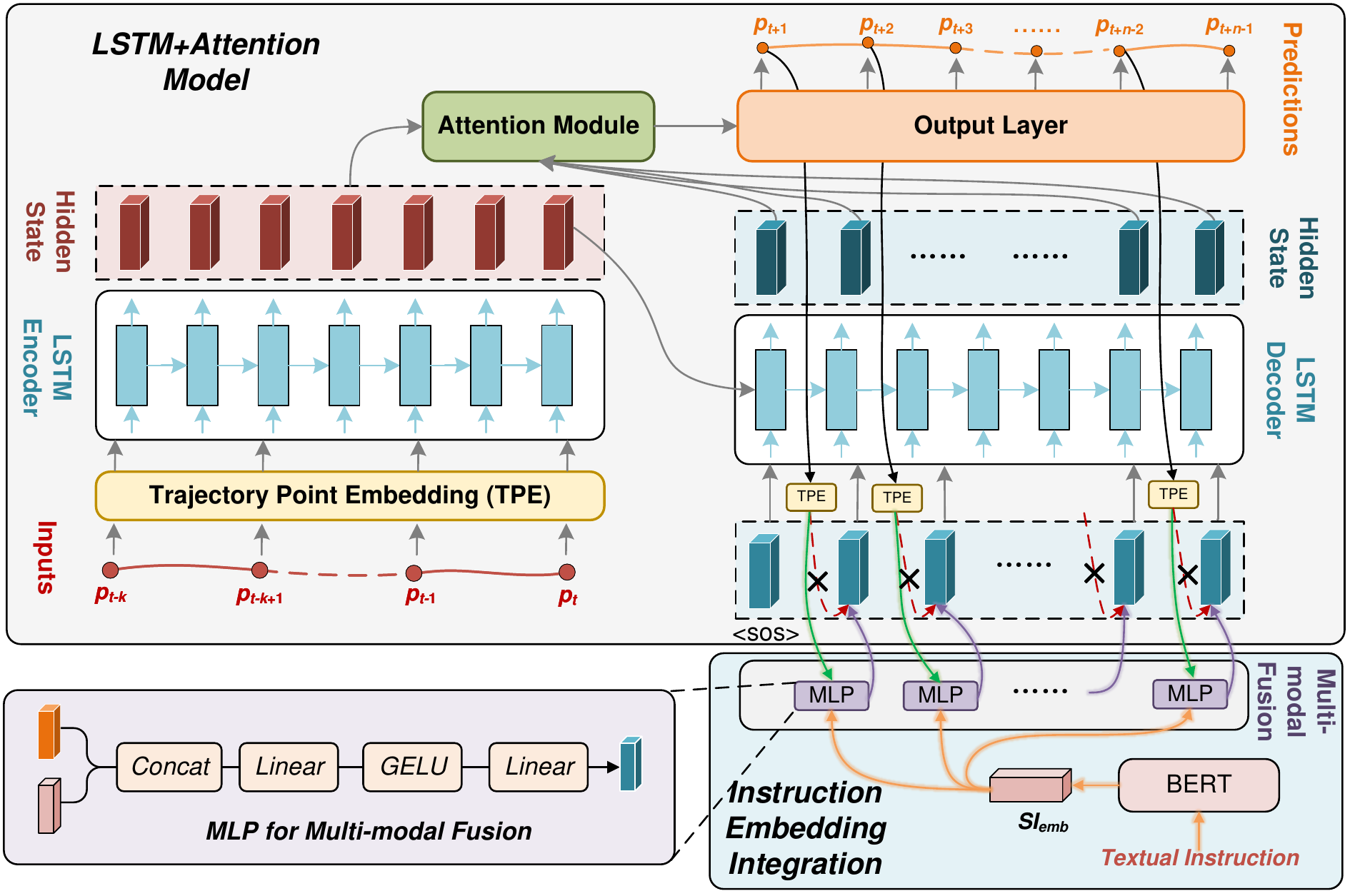}
	\caption{The implementation details of the SIA-FTP framework using LSTM+Attention, in which the LSTM+Attention model is conducted on a vanilla attention-based sequence-to-sequence architecture. 
  In the trajectory-based FTP pre-training stage, the model generates the predictions regressively along the path delineated by the \textit{red dotted line} (output from the first TPE model). 
  In the multi-modal FTP finetuning stage, the predicted trajectory point is embedded by the TPE module and further applied to perform multi-modal fusion in the instruction embedding integration module following the path depicted by the \textit{green line}.
  The $<sos>$ is an initial vector of the Decoder.} 
	\label{fig_lstmatt}
\end{figure}

To validate the generalizability of integrating spoken instruction into the FTP procedure and the proposed three-stage progressive learning paradigm, the LSTM+Attention model is applied to alternate the FlightBERT++ to conduct the SIA-FTP framework. 
The detailed implementation of the SIA-FTP framework using the LSTM+Attention model is depicted in Figure \ref{fig_lstmatt}. 
In this implementation, the multi-modal FTP model is also conducted with the three-stage progressive learning paradigm, i.e., trajectory-based FTP pre-training, intent-oriented instruction embedding learning, and multi-modal FTP finetuning. 

In the experiments, the baseline model of the LSTM+Attention is employed to perform the trajectory-based FTP pre-training.
Specifically, the LSTM+Attention model is conducted on a vanilla attention-based sequence-to-sequence architecture, which is composed of a Trajectory Point Embedding (TPE) module, LSTM Encoder, Attention module, LSTM Decoder, and output layer. 
The TPE module is implemented by a Conv1D layer and is used to project the trajectory points into high-dimensional vectors, as that in the FlightBERT++. 
The output layer is built by a Fully Connected layer, which projects the latent vectors into the predicted trajectory attributes. 
The function of the other modules is similar to the vanilla attention-based LSTM Seq2Seq models, more details can be found in \cite{BahdanauCB14}. 
In the intent-oriented instruction embedding learning stage, similar to the original SIA-FTP framework using FlightBERT++, the BERT-based IID model is applied to generate the instruction embedding $\mathbf{SI}_{emb}$. 

In view of the different model architectures between the LSTM+Attention and FlightBERT++, we redesigned the instruction embedding integration strategy in the multi-modal FTP finetuning stage. 
Compared with the FlightBERT++, the LSTM+Attention model predicts the trajectory points regressively in a step-by-step manner. 
Based on this inference paradigm, we fuse the instruction embedding and the trajectory point (predicted) embedding during the step-by-step decoding process. 
As depicted in Figure \ref{fig_lstmatt}, the instruction embedding is integrated into the FTP process by the proposed MLP-based multi-modal fusion mechanism (the path depicted by the \textit{green line}). 
In this way, the model can be aware of the controlling intent at each inference step, thereby facilitating the generation of more precise predictions.

In this implementation, the Z-Score normalization algorithm is applied to process the value into [0,1] for longitude and latitude attributes, while the other attributes are normalized into [0,1] through the Max-min algorithm. 
The TPE module is constructed using a Conv1D layer with a kernel size of 3 and 512 channels, i.e., the dimension of the trajectory point embedding is 512. 
The LSTM Encoder comprises 4 layers of unidirectional LSTM with 512 neurons, whereas the LSTM Decoder is comprised of 2 layers of unidirectional LSTM. 
The dimension of the instruction embedding $\mathbf{SI}_{emb}$ is 256, which is the same as the original SIA-FTP framework. 
Based on the aforementioned settings, the MLP module receives two inputs, i.e., the instruction embedding and trajectory point embedding, fuses and projects them into a single vector (from 768 to 512 dimensions).
In the multi-modal FTP finetuning stage, the hidden state dimension for linear layers in the MLP is set to 1024 and 512, respectively. 
The model is trained utilizing the Adam optimizer with an initial learning rate of $10^{-4}$. 
The software and hardware environments maintain consistency with the original SIA-FTP framework outlined in the manuscript.

\section{Evaluation of the training data size} \label{data_size}
To further validate the impact of dataset size on model performance, we collected and manually annotated additional data over three days (from February 16 to February 18, 2021), 
yielding 618 (D1), 677 (D2), and 785 (D3) trajectory-instruction pairs, respectively. 
We incrementally added these data to the original training set to finetune the multi-modal FTP model in stage 3, which are approximately 10\% (D1), 20\% (D1/D2), and 33\% (D1/D2/D3) of the original training set in stage 3, and evaluate the original test set. 

\begin{figure}[h]
	\centering
	\includegraphics[width=0.45\textwidth]{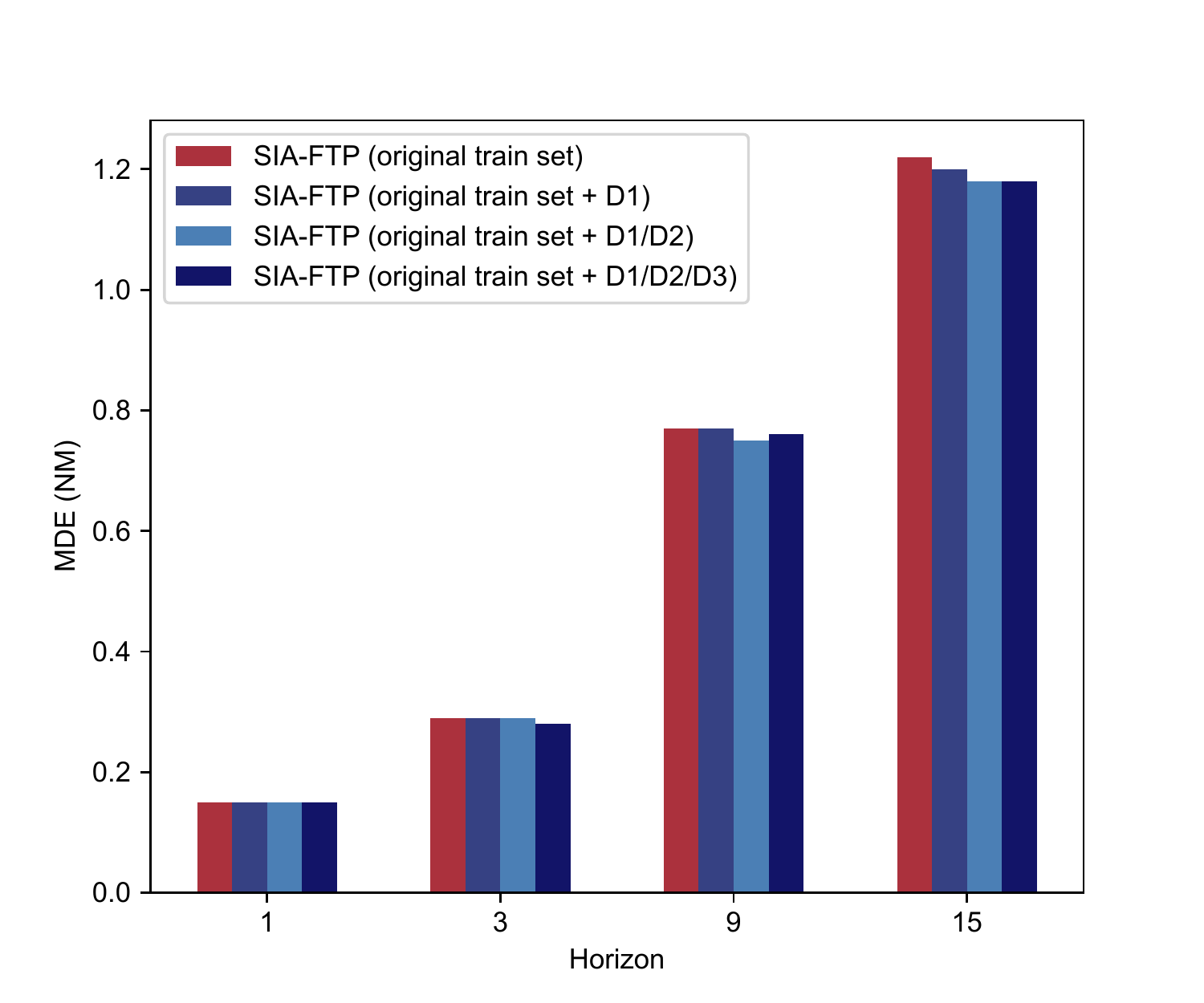}
	\caption{The MDE comparison across the different data sizes of the training set. Source data are provided as a Source Data file.}
	\label{fig_data_size}
\end{figure}

The comparison of the MDE across these training sets is illustrated in Figure \ref{fig_data_size}.
It can be observed from the results that increasing the training data only obtains marginal improvements in model performance. 
Specifically, there is no significant enhancement in prediction performance at a horizon of 15 when the data is increased from 20\% to 33\% (from D1/D2 to D1/D2/D3). 
Meanwhile, a slight decline in performance is noted at a horizon of 9, demonstrating that the model has likely reached convergence and fluctuates in a small range. 
This result can be attributed that the proposed 3-stage training paradigm has ability to reduce the requirements of the paired trajectory-instruction samples, especially the pre-training stage 1 and stage 2-1 benefit to learning the universal data representations from the large-scale unlabeled data samples. 
In summary, although the increasing training samples provide slight performance improvement, compared to the annotation costs of the human resource, developing a non-data-hungry model is a preferred solution to improve the prediction performance.

\end{document}